\RequirePackage{amsmath}
\documentclass[twocolumn,epjc3]{svjour3}
\pdfoutput=1
\RequirePackage{graphicx}
\RequirePackage{hyperref}
\RequirePackage{microtype}
\RequirePackage{cite}
\RequirePackage{slashed}
\RequirePackage{mathptmx}
\RequirePackage[british]{babel}

\newcommand{\alem}{\alpha_\text{em}}
\newcommand{\Delrho}[2][{}]{\Delta\rho_\text{#1}^\text{(#2)}}
\newcommand{\delrho}[2][{}]{\delta\rho_\text{#1}^\text{(#2)}}
\newcommand{\MhO}{m_{h^0}}
\newcommand{\MHp}{m_{H^\pm}}
\newcommand{\MHH}{m_{H^0}}
\newcommand{\MAO}{m_{A^0}}
\newcommand{\dMVsq}[1]{\delta M_{#1}^2}
\newcommand{\dMWsq}{\delta M_W^2}
\newcommand{\dMZsq}{\delta M_Z^2}

\newcommand{\dMHHsq}{\delta m_{H^0}^2}
\newcommand{\dMAOsq}{\delta m_{A^0}^2}
\newcommand{\dMHpsq}{\delta m_{H^\pm}^2}
\newcommand{\dSWsq}{\delta s_W^2}
\newcommand{\dCWsq}{\delta c_W^2}
\newcommand{\Self}[3][\text{}]{\Sigma^{#2}_{#1}\left(#3\right)}
\newcommand{\Sigmaf}[2]{\Sigma_f^{#1}\left(#2\right)}
\newcommand{\SBA}{s_{\beta-\alpha}}
\newcommand{\CBA}{c_{\beta-\alpha}}
\newcommand{\SB}{s_\beta}
\newcommand{\CB}{c_\beta}
\newcommand{\TB}{t_\beta}
\newcommand{\SA}{s_\alpha}
\newcommand{\CA}{c_\alpha}
\newcommand{\STwoB}{s_{2\beta}}

\newcommand{\lfive}{\lambda_5}
\newcommand{\STwoA}{s_{2\alpha}}

\newcommand{\Lag}{\mathcal{L}}
\renewcommand{\Re}{\operatorname{Re}}

\newcommand{\order}[1]{\mathcal{O}\left(#1\right)}
\newcommand{\Matr}{\mathbf{M}}
\newcommand{\Rot}[1]{\mathbf{R}\left(#1\right)}

\newcommand{\Phic}{\Phi^\dagger}
\newcommand{\mat}{\mathcal{M}}

\newcommand{\Tr}[1]{\text{Tr}#1}
\newcommand{\PhiSM}{\Phi_\text{SM}}
\newcommand{\PhiNS}{\Phi_\text{NS}}

\newcommand{\GeV}{{\rm GeV}}

\journalname{Eur. Phys. J. C}

\begin{document}

\title{\boldmath Two-loop corrections to the $\rho$ parameter in Two-Higgs-Doublet models}

\author{Stephan Hessenberger\thanksref{e1,addr1}
        \and
        Wolfgang Hollik\thanksref{e2,addr1} %etc.
}

\thankstext{e1}{e-mail: shessen@mpp.mpg.de}
\thankstext{e2}{e-mail: hollik@mpp.mpg.de}

\institute{Max-Planck-Institut f\"ur Physik (Werner-Heisenberg-Institut), F\"ohringer Ring 6, D--80805 M\"unchen, Germany\label{addr1}
}

\date{Received: date / Accepted: date}
% The correct dates will be entered by the editor

\maketitle

\begin{abstract}
Models with two scalar doublets are among the simplest extensions of
the Standard Model which fulfill the relation $\rho = 1$ at lowest
order for the $\rho$ parameter as favored by experimental data for
electroweak observables allowing only 
small deviations from unity. Such small deviations $\Delta\rho$ originate
exclusively from quantum effects  with special sensitivity to mass splittings 
between different isospin components of fermions and scalars.    
In this paper
the dominant two-loop electroweak corrections to $\Delta\rho$ are calculated
in the $CP$-conserving THDM, resulting from the top-Yukawa coupling and
the self-couplings of the Higgs bosons in the gauge-less limit.
The on-shell renormalization scheme is applied.
With the assumption that one
of the $CP$-even neutral scalars represents the scalar boson observed
by the LHC experiments, with standard properties, the two-loop non-standard
contributions in $\Delta\rho$ can be separated from the standard ones.
These contributions are of particular interest since they 
increase with mass splittings between
non-standard Higgs bosons and can be additionally enhanced by $\tan\beta$
and $\lambda_5$, an additional free coefficient of the Higgs potential,
and can thus modify the one-loop result substantially.
Numerical results are given for the dependence on the various
non-standard parameters, and the influence on the calculation of
electroweak precision observables is discussed.
\end{abstract}

\section{Introduction}
High-precision experiments at electron-positron and hadron colliders 
together with  highly accurate measurements at low energies
have imposed
stringent tests on the Standard Model (SM) and possible extensions.
The experimental accuracy
in the electroweak observables is sensitive to the quantum effects
and requires the highest standards
on the theoretical side as well. 
A sizeable amount of theoretical work has
contributed over more than two decades to a steadily rising
improvement of the SM predictions and also for
specific new physics scenarios like supersymmetric extensions.
The highly accurate measurements and theoretical predictions,
at the level of 0.1\% precision and better, provide unique tests of 
the quantum structure of the SM, which has  been impressively  
confirmed by the discovery of a Higgs particle
by ATLAS~\cite{Aad:2012tfa} and CMS~\cite{Chatrchyan:2012ufa}.
Moreover, it opens the possibility to obtain  indirect informations 
on potential heavy new physics beyond the SM, 
in particular on the not yet sufficiently explored scalar sector.

With the meanwhile very precisely measured Higgs-boson 
mass~\cite{Aad:2015zhl} of
$M_H = 125.09 \pm 0.24 \; \GeV$
the SM input is now completely determined
and the SM predictions for the set of precision observables are unique,
being in overall good agreement with the data.
This improves the sensitivity to physics beyond the SM and
makes constraints on parameters of extended models quite severe.

Models with two scalar doublets in the Higgs sector are among the simplest extensions of
the Standard Model  (a review on theory and phenomenology can be found in~\cite{Branco:2011iw}).
They fulfill the relation $\rho = 1$ at lowest order for the $\rho$ parameter  as favored by experimental data 
for electroweak precision observables allowing only small deviations from unity. 
Such small deviations $\Delta\rho$ naturally originate exclusively from quantum effects in models with Higgs
doublets, with special sensitivity on mass splittings between different isospin components of fermions and scalars.
$\Delta\rho$ can be related to the vector-boson self-energies
and plays the most prominent role in the higher-order calculation
of precision observables, constituting the leading process independent 
loop corrections to accurately measured quantities
like the $W$-$Z$ mass correlation and the effective weak mixing angle
$\sin^2\theta_{\rm eff}$.

The calculation of electroweak precision observables in the general THDM has a  long  
history ~\cite{Bertolini:1985ia,Hollik:1986gg,Hollik:1987fg,Denner:1991ie,Froggatt:1991qw,Chankowski:1999ta,Grimus:2007if,Grimus:2008nb,LopezVal:2012zb}.
Details of the one-loop renormalization of the THDM have been discussed in various 
papers~\cite{Santos:1996vt,Arhrib:2003ph,LopezVal:2009qy,Kanemura:2014bqa,Kanemura:2015mxa,Krause:2016oke,Denner:2016etu},
with the emphasis of the more recent ones on the Higgs sector aiming
at loop-improved predictions for Higgs-boson observables.
The current status of precision observables is given by complete one-loop calculations which can
be augmented in the subset of the SM loop contributions by
incorporating the known higher-order terms of the SM; the non-standard contribution 
is of one-loop order and systematic two-loop calculations have not
been done (ref.~\cite{LopezVal:2012zb} contains some higher-order
terms by means of effective couplings in the one-loop Higgs contributions).  
This is different from the supersymmetric version of the THDM, the MSSM, 
where the non-standard one-loop corrections to precision observables
have been improved by the two-loop contributions to $\Delta\rho$
resulting from the strong and Yukawa interactions
\cite{Djouadi:1996pa,Djouadi:1998sq,Heinemeyer:2002jq,Haestier:2005ja}.
In order to achieve a similar quality of the theoretical predictions also in the general THDM, 
the first step consists in getting the two-loop contributions to $\Delta\rho$ 
from those sectors where the one-loop effects are large, i.e.\ from
the top-Yukawa interaction and the self-interaction of the extended Higgs scalars.    

In this paper we present the leading two-loop corrections to $\Delta\rho$
in the $CP$-conserving THDM which result from the top-Yukawa coupling and
the self-couplings of the Higgs bosons. Technically they are obtained
in the approximation of the gauge-less limit where the electroweak gauge
couplings are set to zero (and thus the gauge-boson masses, but  keeping $M_W/M_Z$ fixed).
With the assumption that one
of the $CP$-even neutral scalars represents the scalar boson observed
by the LHC experiments, with SM properties, the two-loop non-standard
contributions in $\Delta\rho$ can be separated from the SM ones.
These contributions are of particular interest since they involve corrections
proportional to $m_t^4$, or increase with mass splittings between
non-standard Higgs bosons and can be additionally enhanced by $\tan\beta$
and $\lambda_5$, an additional free coefficient of the Higgs potential.

The paper is organized as follows. Section~\ref{Sec:THDM} 
contains the basic features
of the THDM and specifies the notations, and section~\ref{Sec:Outline}
describes the simplifications made for our two-loop calculation.
Aspects of custodial symmetry in the context of the THDM
are considered in section~\ref{Sec:CustSym} which provide a deeper
understanding of the various higher-order contributions to $\Delta\rho$.
The renormalization scheme is specified in section~\ref{Sec:Renormalization}, and
the calculation of $\Delta\rho$ is described in section~\ref{Sec:rhoparameter}.
The appendix contains the Feynman rules for the counterterm vertices and the definitions of the one- and two-loop scalar integrals. 
The numerical analysis is the content of section~\ref{Sec:Results},  
and conclusions are given in section~\ref{Sec:conclusions}.

%\clearpage

\section{The Two-Higgs-Doublet model}
\label{Sec:THDM}

The THDM Higgs sector consists of two complex $SU(2)_L$ doublet scalar fields with hypercharge $Y=1$:

\begin{equation}
\Phi_1=
\begin{pmatrix}
	\phi_1^+ \\
	\phi_1^0 
\end{pmatrix}, \quad
\Phi_2=
\begin{pmatrix}
	\phi_2^+ \\
	\phi_2^0 
\end{pmatrix}.
\label{Eq:Doublets}
\end{equation}
For our calculation we are using the parameterisation of the potential from \cite{gunion:1990} 
\begin{equation}
\begin{aligned}
V=& \lambda_1\left(\Phi_1^\dagger\Phi_1-\frac{\hat{v}_1^2}{2}\right)^2+\lambda_2\left(\Phi_2^\dagger\Phi_2-\frac{\hat{v}_2^2}{2}\right)^2 \\
& +\lambda_3\left[\left(\Phi_1^\dagger\Phi_1-\frac{\hat{v}_1^2}{2}\right)+\left(\Phi_2^\dagger\Phi_2-\frac{\hat{v}_2^2}{2}\right)\right]^2\\
& +\lambda_4\left[\left(\Phi_1^\dagger\Phi_1\cdot\Phi_2^\dagger\Phi_2\right)-\left(\Phi_1^\dagger\Phi_2\cdot\Phi_2^\dagger\Phi_1\right)\right] \\
& +\lambda_5\left[\operatorname{Re}\left(\Phi_1^\dagger\Phi_2\right)-\frac{\hat{v}_1 \hat{v}_2}{2}\right]^2 \\
&+\lambda_6\left[\operatorname{Im}\left(\Phi_1^\dagger\Phi_2\right)\right]^2,
\end{aligned}
\label{Eq:potential}
\end{equation}
in which all the parameters appearing in  the potential are chosen real. 
The form of \eqref{Eq:potential} represents the most general potential involving two doublets and is $CP$-conserving, gauge-invariant, renormalizable and subject to a discrete $Z_2$ symmetry ($\Phi_2\rightarrow - \Phi_2$) which is only softly violated by dimension-two terms. The vacuum expectation values are written as

\begin{equation}
 \langle\Phi_i\rangle=
 \begin{pmatrix}
  0 \\
  \frac{v_i}{\sqrt{2}}
 \end{pmatrix};\quad i=1,2,
\end{equation}
such that the electromagnetic gauge-symmetry $U(1)_\text{em}$ is preserved. The two doublets can be expanded around the vacuum expectation values
\begin{equation}
\Phi_i=
\left(\begin{array}{c}
	\phi_i^+ \\
	\frac{1}{\sqrt{2}}\left(v_i+\eta_i+i\chi_i\right) 
\end{array}\right);\quad i=1,2.
\label{scalarparam}
\end{equation}
Inserting this decomposition into the potential yields for the linear
and quadratic terms the following component form:
\begin{align}
V =& -T_{\eta_1}\eta_1-T_{\eta_2}\eta_2 \notag\\
    &+
	\begin{pmatrix}
		\phi_1^- & \phi_2^-
	\end{pmatrix}
	\hat{\Matr}^\phi
	\begin{pmatrix}
		\phi_1^+ \\
		\phi_2^+
	\end{pmatrix}\notag\\
	 &+\frac{1}{2}
	\begin{pmatrix}
		\chi_1 & \chi_2
	\end{pmatrix}
	\hat{\Matr}^\chi
	\begin{pmatrix}
		\chi_1 \\
		\chi_2
	\end{pmatrix}\notag\\
	&+\frac{1}{2}
	\begin{pmatrix}
		\eta_1 & \eta_2
	\end{pmatrix}
	\hat{\Matr}^\eta
	\begin{pmatrix}
		\eta_1 \\
		\eta_2
	\end{pmatrix}+\dots.
	\label{Eq:Vmass}
\end{align}

The explicit expressions for the tadpoles and the mass matrices in \eqref{Eq:Vmass} can be written in a compact form with the help of the quantities
\begin{eqnarray}
T_{1}=& \lambda _1 \left(v_1^2-\hat{v}_1^2\right),\\
T_{2}=& \lambda _2 \left(v_2^2-\hat{v}_2^2\right),\\
T_{3}=& \lambda _3 \left(v_1^2+v_2^2-\hat{v}_1^2-\hat{v}_2^2\right),\\
T_{5}=&\frac{1}{2} \lambda _5 \left(v_1 v_2-\hat{v}_1 \hat{v}_2\right).
\end{eqnarray}
The tadpoles are given by
\begin{eqnarray}
 T_{\eta_1}&=-T_5 v_2-\left(T_1 + T_3\right) v_1,\label{Eq:eta1tadpole} \\
 T_{\eta_2}&=-T_5 v_1-\left(T_2+T_3\right) v_2.
\label{Eq:eta2tadpole}
 \end{eqnarray}
The mass matrices can be decomposed as
\begin{equation}
  \hat{\Matr}^X=\Matr^X+\Matr^{T};\quad X={\phi,\chi,\eta},
\end{equation}
with
\begin{equation}
 \Matr^T=\begin{pmatrix}
 T_1+T_3 & T_5 \\
 T_5 & T_2+T_3 \\
\end{pmatrix}
\end{equation}
 and
\begin{align}
 \Matr^\eta=&
\begin{pmatrix}
 2 \left(\lambda _1+\lambda _3\right) v_1^2+\frac{v_2^2}{2}  \lambda _5 &  v_1 v_2 \left(2 \lambda _3+\frac{\lambda _5}{2}\right) \\
  v_1 v_2 \left(2 \lambda _3+\frac{\lambda _5}{2}\right) & 2 v_2^2 \left(\lambda _2+\lambda _3\right) +\frac{v_1^2}{2} \lambda _5 \\
\end{pmatrix},\\
\Matr^\chi=&\frac{1}{2}
\begin{pmatrix}
  v_2^2 \lambda _6 & - v_1 v_2 \lambda _6 \\
 - v_1 v_2 \lambda _6 &  v_1^2 \lambda _6 \\
\end{pmatrix},\\
\Matr^\phi=& \frac{1}{2}
\begin{pmatrix}
  v_2^2 \lambda _4 & - v_1 v_2 \lambda _4 \\
 - v_1 v_2 \lambda _4 &  v_1^2 \lambda _4 \\
\end{pmatrix}.
\end{align}
 The requirement that the tadpoles in \eqref{Eq:eta1tadpole} and \eqref{Eq:eta2tadpole} must vanish results in the minimum conditions
\begin{equation}
 \hat{v}_{1,2}=v_{1,2}.
 \label{Eq:Mincondition}
\end{equation}
With the help of these minimum conditions the mass matrices take the form $\Matr^\eta$, $\Matr^\chi$ and $\Matr^\phi$ which have to be diagonalized in order to obtain the physical Higgs states. The unitary transformations
\begin{eqnarray}
\begin{pmatrix}
	G^\pm \\
	H^\pm 
\end{pmatrix}
=&
\Rot{\beta}
\begin{pmatrix}
\phi_1^\pm \\
\phi_2^\pm \\
\end{pmatrix}
\label{Eq:chargrot},\\
\begin{pmatrix}
	G^0 \\
	A^0 
\end{pmatrix}
=&
\Rot{\beta}
\begin{pmatrix}
\chi_1 \\
\chi_2 \\
\end{pmatrix},
\label{Eq:chirot}\\
\begin{pmatrix}
H^0 \\
h^0
\end{pmatrix}
=&
\Rot{\alpha}
\begin{pmatrix}
\eta_1 \\
\eta_2
\end{pmatrix},
\label{Eq:etarot}
\end{eqnarray}
with
\begin{equation}
 \Rot{x}=
 \begin{pmatrix}
\cos x & \sin x \\
-\sin x & \cos x \\ 
\end{pmatrix}
\end{equation}
lead to five physical mass eigenstates: two $CP$-even states $h^0$ and $H^0$, a $CP$-odd state $A^0$ and a charged pair $H^\pm$. $G^0$ and $G^\pm$ are the usual Goldstone bosons associated with the longitudinal modes of the gauge bosons. The mixing angles are determined by 
\begin{equation}
\tan\beta=\frac{v_2}{v_1}
\end{equation}
and
\begin{equation}
\sin2\alpha=\frac{2\Matr^\eta_{12}}{\sqrt{\left(\Matr^\eta_{11}-\Matr^\eta_{22}\right)^2+4(\Matr^\eta_{12})^2}}.
\end{equation}
From now on we will use the short notation $\sin x=s_x$, $\cos x=c_x$ and $\tan x=t_x$ for all the appearances of the mixing angles. 

The kinetic terms of $\Phi_{1,2}$ in the Lagrangian describe the interactions between the scalar fields and the gauge fields and give rise to the gauge-boson masses
\begin{equation}
M_W^2=\frac{g_2^2 (v_1^2+v_2^2)}{4},
\end{equation}
\begin{equation}
M_Z^2=\frac{\left(g_1^2+g_2^2\right)(v_1^2+v_2^2)}{4}.
\end{equation}
$g_1$ is the gauge coupling of $U(1)_Y$ and $g_2$ is the gauge coupling of the $SU(2)_L$. They are also contained in the definition of the electroweak mixing angle $\theta_W$ by
\begin{eqnarray}
 \cos^2\theta_W&=c_W^2=\frac{M_W^2}{M_Z^2},\label{Eq:CW_Definition}\\
 \sin^2\theta_W&=s_W^2=1-c_W^2\label{Eq:SW_Definition}
\end{eqnarray}
and the electric charge
\begin{equation}
 e=\sqrt{4 \pi \alem}=\frac{g_1 g_2}{\sqrt{g_1^2+g_2^2}}
\end{equation}
with the electromagnetic fine structure constant $\alem$.

After electroweak symmetry breaking the combination $v^2=v_1^2+v_2^2$ is fixed by the masses of the gauge bosons. The other seven free parameters of the Higgs potential can be rewritten in terms of the Higgs masses, the mixing angle $\alpha$, the ratio of the vacuum expectation values $\TB$ and the remaining self coupling parameter $\lambda_5$: 
\begin{align}
\lambda_1=&\frac{e^2}{8M_W^2s_W^2}\left[\frac{\CBA\SA}{\SB\CB^2 }m_{h^0}^2
+\frac{\SBA\CA}{\SB\CB^2 }m_{H^0}^2\right]\notag\\
&+\frac{1}{4}\left(1-\TB^2\right)\lambda_5,
\\
\lambda_2=&\frac{e^2}{8M_W^2s_W^2}\left[\frac{\CBA\CA}{\SB^2\CB }m_{h^0}^2
-\frac{\SBA\SA}{\SB^2\CB }m_{H^0}^2\right]\notag\\
&+\frac{1}{4}\left(1-\frac{1}{\TB^2}\right)\lambda_5,
\\
\lambda_3=&-\frac{e^2}{8M_W^2s_W^2}\frac{\STwoA}{\STwoB}\left(m_{h^0}^2-m_{H^0}^2\right)-\frac{\lambda_5}{4},
\\
\lambda_4=&\frac{e^2m^2_{H^\pm}}{2M_W^2s_W^2},
\\
\lambda_6=&\frac{e^2m^2_{A^0}}{2M_W^2s_W^2}.
\end{align}

The terms in \eqref{Eq:Vmass} can be written in the mass eigenstate basis by applying the rotations from \eqref{Eq:chargrot}, \eqref{Eq:chirot} and \eqref{Eq:etarot} before employing the minimum conditions. From the terms linear in the $CP$-even components $\eta_1$ and $\eta_2$ we obtain the tadpoles
\begin{eqnarray}
 T_{h}=&-s_\alpha T_{\eta_1} + c_\alpha T_{\eta_2},\\
 T_{H}=&c_\alpha T_{\eta_1} +s_\alpha T_{\eta_2},
\end{eqnarray}
of the fields $h^0$ and $H^0$. From the matrix $\Matr^T$ in the quadratic terms we obtain mass terms for the Goldstone bosons $G^0$ and $G^\pm$ with
\begin{equation}
 m_{G^0}^2=m_{G^\pm}^2=-\frac{e}{2 M_W s_W} \left(\SBA T_{h}+\CBA T_{H}\right)
 \label{Eq:Goldstonemass}
\end{equation}
and off-diagonal elements in the mass matrices in the mass eigenstate basis. The minimum condition in \eqref{Eq:Mincondition} is then equivalent to the requirement that the tadpole coefficients $T_{\{h^0,H^0\}}$ and all off-diagonal elements in the mass matrices must vanish. An additional consequence is that the Goldstone bosons receive no masses from the Higgs potential.

The couplings between the scalars and the fermions are restricted by the experimental limits on tree-level flavor changing neutral currents (FCNCs). It has been shown in \cite{Glashow:1976nt} and \cite{Paschos:1976ay} that a necessary and sufficient condition to avoid the FCNCs by neutral Higgs exchange at tree level is that not more than one of the doublets couples to fermions of a given charge. This has lead to four main models which are discussed in the literature

\begin{itemize}
 \item type-I: all leptons and quarks couple only to the doublet $\Phi_2$;
 \item type-II: the up-type quarks couple to the doublet $\Phi_2$, while all the down-type quarks and leptons couple to the doublet $\Phi_1$;
 \item type-X or lepton specific model: all quarks couple to $\Phi_2$ and all leptons couple to $\Phi_1$;
 \item type-Y or flipped model: the up-type quarks and leptons couple to the doublet $\Phi_2$ while the down-type quarks couple only to $\Phi_1$.
\end{itemize}

In the top-Yukawa approximation all the Yukawa couplings besides the one of the top quark, are neglected.  Since the top-Yukawa coupling is given in all models by the interaction of the up-type quarks with the doublet $\Phi_2$, our result is valid in each of the four models. Since we are assuming a diagonal CKM matrix the top-Yukawa term of the Lagrangian takes the form 

\begin{align}
 \Lag_{Y,t}=&-m_t\overline{\psi}_t\psi_t-\frac{e m_t}{2 M_W s_W}\overline{\psi}_t\psi_t\left(\frac{\SA}{\SB}H^0+\frac{\CA}{\SB}h^0\right)\notag\\
  &+i\frac{e m_t}{2 M_W s_W}\overline{\psi}_t\gamma_5\psi_t\left(G^0+\frac{1}{\TB}A^0\right)\notag\\
  &+\frac{e m_t}{\sqrt{2} M_W s_W}\overline{\psi_t}\omega_-\psi_b\left(G^+ +\frac{1}{\TB}H^+\right)+h.c.
  \label{Eq:topLagrangian}
\end{align}
where $m_t$ is the mass of the top quark, $\psi_{t,b}$ are the Dirac spinors of the top and bottom quarks and $\omega_-=(1-\gamma_5)/2$ is the projector on the left-handed spinor states. 

Models with a more general structure for the Higgs-fermion interactions are usually refered to as type-III models \cite{Hou:1991un,Chang:1993kw,Atwood:1996vj} and allow couplings of all the SM fermions to both Higgs doublets. The more general Higgs-fermion couplings are then strongly restricted by the absence of FCNCs.

\section{Approximations and outline of the calculation}
\label{Sec:Outline}

In order to evaluate the leading two-loop contributions from the
Yukawa sector and from the Higgs self-interactions
a number of approximations can be made.

\subsection{Gauge-less limit and top-Yukawa approximation}
Since our focus is on the corrections to $\Delta\rho$ originating from
the top-Yukawa and the Higgs self-couplings in the THDM we neglect all
other couplings. This means that we work in the gauge-less limit 
(as done in \cite{Haestier:2005ja} for the MSSM) 
in which the electroweak gauge couplings $g_{1,2}$ are put to zero and
thus the gauge-boson masses are also equal to zero
\begin{equation}
M_W^2=\frac{g_2^2 v^2}{4}\rightarrow 0,\quad
M_Z^2=\frac{\left(g_1^2+g_2^2\right)v^2}{4}\rightarrow 0 \, ,
\end{equation}
while their ratio in $c_W$ and $s_W$ stays constant. 
Moreover, the masses of the Goldstone bosons  are zero,
\begin{equation}
 m_{G^0}=m_{G^\pm}=0 ,
\end{equation}
in the gauge-less limit.

In addition we are using the top-Yukawa approximation in which all the
fermion masses with the exception of the top-quark mass are
neglected. Especially for the bottom quark, which appears in some of
the diagrams for the $\order{\alpha_t^2}$ contributions, we set $m_b=0$.
%\begin{equation}
%m_b=0.
%\end{equation}
%This is justified as long as we do not consider
%large values of $\tan\beta$ when Yukawa-interaction effects from  $b$-quarks  
%can become enhanced.

Differently from the top-Yukawa coupling, which is universal in
  all of the four models, the Yukawa coupling of the bottom quark is
  model specific. In models of type-I and type-X, the bottom- and
  top-Yukawa interactions have the same structure, and the
  additional contributions to $\Delta\rho$  from the $b$~quark are negligible due to the
  small value of $m_b$. In models of type-II or type-Y, the $b$-Yukawa
  coupling can be enhanced by $\TB$, and the top-Yukawa approximation 
  is justified in these models as long as we do not consider large
  values of $\TB$. For large $\TB$ values additional constraints from
  flavor physics would have to be taken into account as well.

\subsection{The alignment limit}\label{subsec:Align}

Due to the fact that a scalar particle with a mass of approximately
125 GeV has been observed at the LHC~\cite{Aad:2012tfa,Chatrchyan:2012ufa} 
we can identify one of the $CP$-even scalars with the observed resonance.
Choosing $h^0$ (without loss of generality) corresponds to setting
\begin{equation}
 \MhO=125\text{ GeV}.
 \label{Eq:Mh0value}
\end{equation}
Furthermore the analysis of the Higgs couplings by \mbox{ATLAS} \cite{Aad:2015gba} and CMS \cite{Khachatryan:2014jba} indicate no significant deviations from the couplings of the Higgs boson in the SM. Therefore we choose to work in the alignment limit \cite{Dev:2014yca,Bernon:2015qea}, in which 
the angles are correlated via
\begin{equation}
 \alpha=\beta-\frac{\pi}{2}
 \label{Eq:Alignalpha}
\end{equation}
and the couplings of $h^0$ to the vector bosons and fermions are identical to the corresponding couplings of the Higgs boson in the SM. In this limit the $CP$-even Higgs states are obtained by
\begin{equation}
\begin{pmatrix}
H^0 \\
h^0
\end{pmatrix}
=
\begin{pmatrix}
\SB & -\CB \\
\CB & \SB \\
\end{pmatrix}
\begin{pmatrix}
\eta_1 \\
\eta_2
\end{pmatrix}
\end{equation}
and the two doublets can be rewritten as
\begin{eqnarray}
\Phi_1&=\CB\Phi_\text{SM}-\SB\Phi_\text{NS},\\
\Phi_2&=\SB\Phi_\text{SM}+\CB\Phi_\text{NS}
\end{eqnarray}
with
\begin{eqnarray}
\Phi_\text{SM}
&=
\begin{pmatrix}
G^+\\
\frac{1}{\sqrt{2}}\left(v+h^0+iG^0\right)
\end{pmatrix},\label{Eq:SMdoublet}\\
\Phi_\text{NS}
&=
\begin{pmatrix}
H^+\\
\frac{1}{\sqrt{2}}\left(-H^0+iA^0\right)
\end{pmatrix}.
\label{Eq:NSdoublet}
\end{eqnarray}
Moreover, the relations for $\lambda_1$, $\lambda_2$ and $\lambda_3$ simplify to

\begin{eqnarray}
\lambda_1=&\frac{e^2}{8M_W^2s_W^2}\frac{m_{H^0}^2}{\CB^2 }
+\frac{1}{4}\left(1-\TB^2\right)\lambda_5
%\notag\\
%=&\frac{e^2}{8M_W^2s_W^2}m_{H^0}^2\left(1+\TB^2\right)
%+\frac{1}{4}\left(1-\TB^2\right)\lambda_5 ,
\\
\lambda_2=&\frac{e^2}{8M_W^2s_W^2}
\frac{m_{H^0}^2}{\SB^2}+\frac{1}{4}\left(1-\frac{1}{\TB^2}\right)\lambda_5
%\notag\\
%  &=\frac{e^2}{8M_W^2s_W^2}
%m_{H^0}^2\left(1+\frac{1}{\TB^2}\right)+\frac{1}{4}\left(1-\frac{1}{\TB^2}\right)\lambda_5 ,
\\
\lambda_3=&\frac{e^2}{8M_W^2s_W^2}\left(m_{h^0}^2-m_{H^0}^2\right)-\frac{\lambda_5}{4}.
\end{eqnarray}
The potential can be rewritten in terms of the doublets given in \eqref{Eq:SMdoublet} and \eqref{Eq:NSdoublet}. For the classification of the different contributions to $\Delta\rho$ we split it in the four parts

\begin{align}
 V=&V_{I}+V_{II}+V_{III}+V_{IV};\label{Eq:PotAlign}\\
 V_{I}=&\frac{m_{h^0}^2 }{2 v^2}\left(\Phi_{\text{SM}}^{\dagger } \Phi _{\text{SM}}\right){}^2+\frac{1}{2}m_{h^0}^2\left(\frac{1}{4} v^2 - \Phi_{\text{SM}}^{\dagger } \Phi_{\text{SM}}\right),\label{Eq:V_I}\\
 V_{II}=&\frac{1}{2 v^2}\left(\Phi_{\text{NS}}^{\dagger } \Phi _{\text{NS}}\right){}^2 \left(m_{h^0}^2+\frac{4 m_{H^0}^2}{t_{2 \beta }^2}-\frac{2 \lambda _5 v^2}{t_{2 \beta }^2}\right)
 \notag\\
 &+\frac{1}{2} \left(\lambda _5 v^2-m_{h^0}^2\right) \left(\Phi_{\text{NS}}^{\dagger } \Phi _{\text{NS}}\right),\label{Eq:V_II}\\
 V_{III}=&\frac{\left(m_{A^0}^2-2 \MHp^2+m_{H^0}^2\right)}{v^2}   
    \left(\Phi_{\text{SM}}^{\dagger } \Phi _{\text{NS}}\cdot\Phi_{\text{NS}}^{\dagger } \Phi_{\text{SM}}\right)  \notag\\
	&+\frac{\left(m_{H^0}^2-m_{A^0}^2\right)}{2 v^2}\left( \left(\Phi_{\text{NS}}^{\dagger } \Phi _{\text{SM}}\right){}^2+\left(\Phi_{\text{SM}}^{\dagger } \Phi _{\text{NS}}\right){}^2\right)\notag\\
	&+ \left(\frac{2 \MHp^2+m_{h^0}^2}{v^2}-\lambda _5\right)\left(\Phi_{\text{NS}}^{\dagger } \Phi _{\text{NS}} \cdot \Phi_{\text{SM}}^{\dagger } \Phi _{\text{SM}}\right),\label{Eq:V_III}\\
 V_{IV}=&\frac{1}{t_{2 \beta }}\left(\frac{2 m_{H^0}^2}{v^2}-\lambda _5\right)\cdot\notag\\
	  &\left( \Phi_{\text{NS}}^{\dagger } \Phi _{\text{NS}}\cdot \Phi_{\text{NS}}^{\dagger } \Phi _{\text{SM}}+\Phi_{\text{NS}}^{\dagger } \Phi _{\text{NS}}\cdot \Phi_{\text{SM}}^{\dagger } \Phi_{\text{NS}}\right)\label{Eq:V_IV}.
\end{align}

Imposing \eqref{Eq:Alignalpha} on the top-Yukawa interaction given in
\eqref{Eq:topLagrangian} one finds that the resulting coupling between the SM-like scalar $h^0$ and the top quark is identical to the top-Yukawa coupling in the SM, while the couplings to the non-standard Higgs states $A^0$, $H^0$ and $H^\pm$ receive an additional factor of $\TB^{-1}$.
The various types I, II, X, Y of THDMs coincide within the approximations made in this paper.

\subsection{Outline of the calculation}

All needed diagrams and amplitudes are generated with the help of the Mathematica package \texttt{FeynArts} \cite{Hahn:2000kx}. The
evaluation of the one-loop amplitudes and the calculation of the renormalization constants is done with the help of
the package \texttt{FormCalc} \cite{Hahn:1998yk}, which is also employed to generate a Fortran expression of the result. 
In the numerical analysis of the one-loop result the integrals are evaluated with the program \texttt{LoopTools} \cite{Hahn:1998yk}.

The package \texttt{TwoCalc} \cite{Weiglein:1993hd,Weiglein:1992rj} is applied to deal with the Lorentz and Dirac algebra of the two-loop
amplitudes and to reduce the tensor integrals to scalar integrals. In the gauge-less limit the external momenta of all the two-loop diagrams
are equal to zero and the result depends only on the one-loop functions $A_0$ and $B_0$ (see appendix~\ref{App:1LIntegrals}) and on the
two-loop function $T_{134}$ (see appendix~\ref{App:2LIntegrals}) for which analytic expressions are known~\cite{Davydychev:1992mt,Berends:1994ed} 
and Fortran functions are encoded in the program \texttt{FeynHiggs} \cite{Heinemeyer:1998yj,Hahn:2009zz}.
For the automation of the calculation and the implementation of the result in Fortran, the techniques from \cite{Hahn:2015gaa} are
employed.

\section{Custodial symmetry in the SM and the THDM}
\label{Sec:CustSym}

The custodial symmetry is an approximate global \linebreak${SU(2)_L\times SU(2)_R}$ symmetry of the SM which is responsible for the tree-level value of the $\rho$ parameter \cite{Weinberg:1979bn,Susskind:1978ms,Sikivie:1980hm}. Since the Higgs potential respects the remaining $SU(2)_{L+R}$ after electroweak symmetry breaking the $\rho$ parameter is protected from large radiative corrections in the Higgs mass. In the gauge interaction the custodial symmetry is only approximate since it is broken by the hypercharge coupling $g_1$. Moreover, the custodial symmetry is broken by the Yukawa interaction which leads to large corrections to the $\rho$ parameter for large mass differences between quarks in the same doublet \cite{Veltman:1977kh,Chanowitz:1978mv,Chanowitz:1978uj}. A detailed review can be found for example in \cite{Willenbrock:2004hu}.

\subsection{Custodial symmetry in the SM}

As already mentioned the custodial symmetry is a global symmetry of the potential 

\begin{equation}
 V_\text{SM}\left(\Phi\right)=-\mu^2\Phic\Phi+\lambda\left(\Phic\Phi\right)^2,
\end{equation}
with the complex doublet
\begin{equation}
 \Phi=\begin{pmatrix}
       \phi^+ \\
       \phi^0
      \end{pmatrix}.
\end{equation}
To make the symmetry apparent, it is useful to introduce the complex matrix field
\begin{equation}
 \mat=\left(\widetilde{\Phi}|\Phi\right)=
 \begin{pmatrix}
   \phi^{0\ast} & \phi^+ \\
   -\phi^- & \phi^0
 \end{pmatrix}
 \label{Eq:SMmatrix}
\end{equation}
where
\begin{equation}
 \widetilde{\Phi}=i\sigma_2\Phi^\ast=
 \begin{pmatrix}
  0 & 1 \\
  -1 & 0
 \end{pmatrix}
 \begin{pmatrix}
  \phi^- \\
  \phi^{0\ast}
 \end{pmatrix}.
\end{equation}
With this matrix field the potential can be expressed by
\begin{equation}
 V_\text{SM}\left(\mat\right)=-\mu^2\frac{1}{2}\Tr{\mat^\dagger\mat}+\lambda\left(\frac{1}{2}\Tr{\mat^\dagger\mat}\right)^2.
\end{equation}
In addition to the global version of the $SU(2)_L$ gauge symmetry, which transforms $\mat$ according to
\begin{equation}
\mat \rightarrow L \mat
\end{equation}
the potential is  also invariant for $SU(2)_R$ transformations of the form
\begin{equation}
\mat \rightarrow \mat R^\dagger.
\end{equation}
While after electroweak symmetry breaking the vaccum expectation value 
\begin{equation}
 \langle \mat \rangle = \frac{1}{2}  
 \begin{pmatrix}
  v & 0 \\
  0 & v 
 \end{pmatrix}
\end{equation}
 breaks both symmetries 
\begin{equation}
 L \langle \mat \rangle \neq \langle \mat \rangle;\quad \langle \mat \rangle R^\dagger \neq \langle \mat \rangle,
\end{equation}
the potential is still invariant under the subgroup  \linebreak$SU(2)_{L+R}$ of simultaneous $SU(2)_L$ and $SU(2)_R$ transformations with $L=R$, since
\begin{equation}
 L \langle \mat \rangle L^\dagger = \langle \mat \rangle.
\end{equation}

However the custodial symmetry is not an exact symmetry of the SM. It is broken by the hypercharge coupling $g_1$ in the kinetic term of the Higgs Lagrangian which can be written with the matrix field $\mat$ as

\begin{equation}
 \frac{1}{2}\Tr{\left(D_\mu\mat\right)^\dagger \left(D^\mu\mat\right)}
\end{equation}
with the covariant derivative
\begin{equation}
 D_\mu\mat=\left(\partial_\mu\mat+i\frac{g_2}{2}\vec{\sigma}\cdot \vec{W}_\mu \mat-i\frac{g_1}{2}B_\mu\mat\sigma_3\right).
\end{equation}
When neglecting $g_1$ the kinetic term is invariant under the custodial symmetry since $\vec{W_\mu}$ transforms as a triplet under the global $SU(2)_L$,
\begin{equation}
 \vec{\sigma}\cdot\vec{W}_\mu\rightarrow L\vec{\sigma}\cdot\vec{W}_\mu L^\dagger.
\end{equation}

\subsection{Custodial symmetry in the THDM}

A scalar potential with two doublets leads to additional terms which can violate the custodial symmetry. A lot of work has been dedicated to investigations of how the custodial symmetry can be restored in the THDM \cite{Haber:1992py,Pomarol:1993mu,Gerard:2007kn,Grzadkowski:2010dj,Haber:2010bw,Nishi:2011gc}, since there are several possibilities to implement the ${SU(2)_L\times SU(2)_R}$ transformations for two doublets. One way is to introduce matrices similar to \eqref{Eq:SMmatrix} for the two original doublets in \eqref{Eq:Doublets}. The potential is then custodial invariant for $\MHp=\MAO$ \cite{Haber:1992py,Pomarol:1993mu}. Different implementations of the custodial transformations were found in \cite{Pomarol:1993mu,Gerard:2007kn}; these require $\MHp=\MHH$ in order to obtain a custodial-symmetric potential. However, as shown by \cite{Grzadkowski:2010dj,Haber:2010bw,Nishi:2011gc} these different implementations of the $SU(2)_L \times SU(2)_R$ transformations are dependent on the selected basis of the two doublets and can be related to each other by a unitary change of the basis. Since the two doublets have the same quantum numbers, such a change of basis maintains the gauge interaction but modifies the form of the potential and the Yukawa interaction. 

We will demonstrate how the custodial symmetry can be imposed on the potential for the basis of $\PhiSM$ and $\PhiNS$ as defined in \eqref{Eq:SMdoublet} and \eqref{Eq:NSdoublet}.
This choice of basis corresponds to the so-called Higgs basis as defined for example in \cite{Botella:1994cs,Davidson:2005cw} in which only one of the doublets has a non-vanishing vacuum expectation value in its neutral component. Note that the definition of the Higgs basis is only specified up to a rephasing of the second doublet. As explained in \cite{Haber:2010bw}, the only two possible definitions for matrix fields
which preserve the custodial $SU(2)_{L+R}$ after electroweak symmetry breaking are 
\begin{equation}
 \mat_\text{SM}
 =\begin{pmatrix}
   \widetilde{\Phi}_\text{SM}|\Phi_\text{SM}
  \end{pmatrix}
\end{equation}
and
\begin{equation}
 \mat_\text{NS}
 =
 \begin{pmatrix}
  \widetilde{\Phi}_\text{NS} | \Phi_\text{NS} 
 \end{pmatrix}.
\end{equation}
Following \cite{Gerard:2007kn,Haber:2010bw} we write the transformations under the $SU(2)_L\times SU(2)_R$ as 
\begin{equation}
 \mat_\text{SM}\rightarrow L\mat_\text{SM}R^\dagger,\quad
 \mat_\text{NS}\rightarrow L\mat_{\text{NS}}R^{\prime\dagger},
 \label{Eq:Cust_Trans}
\end{equation}
with $L\in SU(2)_L$ and $R,R^\prime\in SU(2)_R$. Since both doublets transform in the same way under the weak $SU(2)_L$ gauge transformations, they have the same transformation matrix $L$ in \eqref{Eq:Cust_Trans}. The same requirement does not hold for transformations under $SU(2)_R$. As explained in \cite{Gerard:2007kn,Haber:2010bw,Nishi:2011gc}, the matrices $R$ and $R^\prime$ are only related by the fact that the doublets $\PhiSM$ and $\PhiNS$ have the same hypercharge and that the $U(1)_Y$ is a subgroup of the $SU(2)_R$. When writing $R=\exp{\left(i\theta n^a T^a_R\right)}$ in terms of an unit vector $n^a$ and the generators $T^a_R=\sigma^a/2$ ($a=1,2,3$), the hypercharge operator for the matrix fields is 
\begin{equation}
 Y=\text{diag}(-1,1)=2T^3_R.
\end{equation}
In order to obtain the same hypercharge transformations for $\mat_\text{SM}$ and $\mat_\text{NS}$ the matrices $R$ and $R^\prime$ are related by
\begin{equation}
R=X^{-1} R^\prime X,
\end{equation}
with 
\begin{equation}
 X\exp{(i\theta Y)}X^{-1}=\exp{(i\theta Y)}.
\end{equation}
This requires the matrix $X$ to have the form
\begin{equation}
 X=\begin{pmatrix}
    e^{-i\chi} & 0 \\
    0 & e^{i\chi}
   \end{pmatrix}, \quad 0\leq\chi\leq 2\pi.
\end{equation}
A scalar potential is invariant under the transformations in \eqref{Eq:Cust_Trans} if it contains only the invariant combinations 
\begin{eqnarray}
 &\Tr{\mat_\text{SM}^\dagger\mat_\text{SM}}=2\Phi^\dagger_\text{SM}\Phi_\text{SM},\\
 &\Tr{\mat_\text{NS}^\dagger\mat_\text{NS}}=2\Phi^\dagger_\text{NS}\Phi_\text{NS},
 \end{eqnarray}
 and
 \begin{equation}
 \Tr{\mat_\text{SM}^\dagger\mat_\text{NS}X}=e^{-i\chi}\PhiNS^\dagger\PhiSM+e^{i\chi}\PhiSM^\dagger\PhiNS.
\end{equation}
The parts $V_{I}$ and $V_\text{II}$ of the potential in \eqref{Eq:PotAlign} are clearly custodial invariant. The parts $V_\text{III}$ and $V_{IV}$ are in general not invariant under the transformations in \eqref{Eq:Cust_Trans}. In order to restore the custodial symmetry the parameters have to be adjusted depending on the value of $\chi$. Since we assumed a $CP$ conserving potential with real parameters this is only possible for $\chi=0$ and $\chi=\pi/2$, as we will show in the following.

\subsubsection{\boldmath Custodial symmetry for \texorpdfstring{$\chi=0$}{chi=0}}
\label{Sec:CustTrans1}

For $\chi=0$, we have $R=R^\prime$ and therefore
\begin{eqnarray}
 \mat_\text{SM}&\rightarrow L\mat_\text{SM}R^\dagger,\\
 \mat_\text{NS}&\rightarrow L\mat_\text{NS}R^\dagger.
\end{eqnarray}
This leads to the invariant quantity
\begin{align}
 \Tr{\mat^\dagger_\text{SM}\mat_\text{NS}X}&=\Tr{\mat^\dagger_\text{SM}\mat_\text{NS}}\notag\\
 &=\Phi_\text{NS}^\dagger\Phi_\text{SM}+\Phi_\text{SM}^\dagger\Phi_\text{NS}.
\end{align}
The part $V_{IV}$ from the potential in \eqref{Eq:PotAlign} is invariant under this custodial transformation since it can be written as follows:
\begin{equation}
\begin{aligned}
 V_{IV}=&\frac{1}{2t_{2 \beta }}\left(\frac{2 m_{H^0}^2}{v^2}-\lambda _5\right)\cdot\\
  &\cdot\Tr{\mat_\text{NS}^\dagger\mat_\text{NS}}\;\Tr{\mat^\dagger_\text{SM}\mat_\text{NS}}
 \end{aligned}
\end{equation}
If we set $\MAO=\MHp$ we can also write $V_\text{III}$ in terms of the invariant quantities,
\begin{align}
 V_{III}&\xrightarrow{\MAO=\MHp}& \notag\\
  &\frac{\MHH^2-\MHp^2}{2v^2} \, \left(\Tr{\mat_\text{SM}^\dagger\mat_\text{NS}}\right)^2\notag\\
  &+\left(\frac{2\MHp^2+\MhO^2}{v^2}-\lambda_5\right)\cdot\notag\\
  &\cdot\frac{1}{4}\Tr{\mat_\text{SM}^\dagger\mat_\text{SM}}\;\Tr{\mat_\text{NS}^\dagger\mat_\text{NS}}
\end{align}
Consequently custodial invariance in the potential can be restored for $\MAO=\MHp$.

\subsubsection{\boldmath Custodial symmetry for \texorpdfstring{$\chi=\frac{\pi}{2}$}{chi=pi/2}}
\label{Sec:CustTrans2}

For $\chi=\frac{\pi}{2}$ we have

\begin{equation}
 X=
 \begin{pmatrix}
  -i & 0\\
  0 & i \\
 \end{pmatrix}
\end{equation}
and
\begin{equation}
 \Tr{\mat_\text{SM}^\dagger\mat_\text{NS}X}=-i\Phi_\text{NS}^\dagger\Phi_\text{SM}+i\Phi_\text{SM}^\dagger\Phi_\text{NS}.
 \label{Eq:Invquant}
\end{equation}
Invariance of $V_\text{III}$ under this custodial transformation is obtained for $\MHH^2=\MHp^2$:
\begin{align}
 V_\text{III}&\xrightarrow{\MHH=\MHp} & \notag\\
 &\frac{\MAO^2-\MHp^2}{2v^2} \, \left(\Tr{\mat_\text{SM}^\dagger\mat_\text{NS}X}\right)^2\notag\\
  &+\left(\frac{2\MHp^2+\MhO^2}{v^2}-\lambda_5\right)\cdot\notag\\
  &\cdot\frac{1}{4}\Tr{\mat_\text{SM}^\dagger\mat_\text{SM}}\;\Tr{\mat_\text{NS}^\dagger\mat_\text{NS}}.
\end{align}
However, the part $V_{IV}$ in the potential cannot be written in
terms of the invariant quantity specified in
\eqref{Eq:Invquant}. Consequently, it has to vanish in the case of a
potential invariant under this custodial transformation. 
This can be achieved by setting
\begin{equation}
 \frac{2\MHH^2}{v^2}=\lfive
\end{equation}
or
\begin{equation}
 \TB=1.
\end{equation}

\section{Renormalization scheme}
\label{Sec:Renormalization}

For our calculation we are using the on-shell renormalization scheme
with the conventions from \cite{Denner:1991kt} in which the masses and
couplings are related to physical parameters. For the renormalization
of the Higgs sector the parameters in the Higgs potential can be
replaced by bare parameters $\hat{v}_{i,0}$ and $\lambda_{i,0}$. Also
the vacuum expectation values $v_1$ and $v_2$ are renormalized in
order to correct for shifts in the minimum of the Higgs potential through
radiative corrections. The resulting renormalization constants can be
translated into counterterms for the masses and mixing angles and for
the tadpoles of $h^0$ and $H^0$. For the subloop renormalization
in the two-loop self-energies we need the following parameters and counterterms:
\begin{align}
M^2_{W,0} & =M_W^2+\dMWsq,\label{Eq:MWbare}\\
M^2_{Z,0} & =M_Z^2+\dMZsq,\label{Eq:MZbare}\\
m_{f,0} & =m_f+\delta m_f, \label{Eq:dMfren}\\
m_{S,0}^2 & = m_S^2 +\delta m_S^2;\quad(S=h^0,H^0,A^0,H^\pm) \label{Eq:dMSren},\\
T_{h,0} & =T_{h}+\delta T_{h},\\
T_{H,0} & =T_{H}+\delta T_{H}.
\end{align}
The tadpole counterterms are fixed such that they cancel all the tadpole diagrams of $h^0$ and $H^0$. 
The resulting renormalization conditions are given by 
\begin{align}
 \delta T_{h}&=-T^{(1)}_{h},\\
 \delta T_{H}&=-T^{(1)}_{H},
\end{align}
where $T^{(1)}_{h,H}$ denote the sum of the respective one-loop Higgs tadpole graphs.
The tadpole counterterms determine the mass counterterms for the Goldstone bosons,
\begin{equation}
\begin{aligned}
 \delta m_{G^0}^2&=\delta m_{G^\pm}^2\\
  &=-\frac{e}{2 M_W s_W} \left(\SBA  \delta T_{h}+\CBA \delta T_{H}\right) ,
 \end{aligned}
\end{equation}
 following from \eqref{Eq:Goldstonemass}. In the alignment limit this simplifies to
\begin{equation}
 \delta m_{G^0}^2=\delta m_{G^\pm}^2=-\frac{e}{2 M_W s_W}  \delta T_{h}.
\end{equation}
In the on-shell scheme mass renormalization is done by the requirement that the renormalized masses are equal to the pole masses,
defined by the real part of the poles of the corresponding propagators. 
Therefore, the mass counterterms 
have to absorb the corrections from the self-energies.  In terms of the gauge-boson self-energies ($V=W,Z$) 
\begin{multline}
\Sigma_{V}^{\mu\nu}\left(p^2\right)=\left(g^{\mu\nu}-\frac{p^\mu p^\nu}{p^2}\right)\Sigma_{V,T}\left(p^2\right)\\
  +\frac{p^\mu p^\nu}{p^2}\Sigma_{V,L}\left(p^2\right) ,
\label{transverseselfenergy}
\end{multline}
the fermion self-energies
\begin{equation}
\Sigma_f\left(p^2\right)=\slashed p \omega_-\Sigma^L_f\left(p^2\right)+\slashed p \omega_+ \Sigma^R_f\left(p^2\right)+m_f\Sigma^S_f\left(p^2\right),
\label{fermionselfdec}
\end{equation}
and the scalar self-energies $\Sigma_S(p^2)$, 
the on-shell renormalization conditions yield the mass counterterms 
%(see also~\cite{Denner:1991kt})
\begin{align}
\dMWsq=&\Re\Self[W,T]{(1)}{M_W^2},\\
\dMZsq=&\Re\Self[Z,T]{(1)}{M_Z^2},\\
\delta m_f=& \frac{m_f}{2}\big[\Re\Sigmaf{L}{m_f^2}+\Re\Sigmaf{R}{m_f^2}\notag\\
  &\hphantom{\frac{m_f}{2}\big[}+2\Re\Sigmaf{S}{m_f^2}\big],\\
\delta m_S^2 =&\Re\Self[S]{}{m_S^2}\quad (S=h^0,H^0,A^0,H^\pm).
\end{align}
The upper index of the gauge-boson self-energies  indicates the loop order, since we need also the two-loop contribution to the gauge-boson
self-energies in the calculation of $\Delta\rho$. For all the other
quantities,  one-loop renormalization is sufficient and we drop the
loop index. Furthermore, we will write $\Sigma_{V}\equiv\Sigma_{V,T}$ ($V=W,Z$) 
for the transverse part of the gauge-boson self-energies.

In the on-shell scheme the definition of the electroweak mixing angle by \eqref{Eq:CW_Definition} and \eqref{Eq:SW_Definition} is valid to all orders in perturbation theory.
Inserting the bare masses from \eqref{Eq:MWbare} and \eqref{Eq:MZbare} yields
\begin{equation}
 s_{W,0}^2=1-c_{W,0}^2=1-\frac{M_{W,0}^2}{M_{Z,0}^2}
 \label{Eq:SWbare}
\end{equation}
 and expanding the ratio of the bare masses up to one-loop order leads to the counterterm
\begin{equation}
\frac{\dSWsq}{s_W^2}=-\frac{c_W^2}{s_W^2}\frac{\dCWsq}{c_W^2}=\frac{c_W^2}{s_W^2}\left(\frac{\dMZsq}{M_Z^2}-\frac{\dMWsq}{M_W^2}\right) .
\end{equation}
In the gauge-less limit the ratios $\dMVsq{V}/M_V^2$ have remaining
contributions, since the gauge couplings of $\order{g_{1,2}^2}$ in the
self-energies cancel with those contained in the gauge-boson
masses. The resulting one-loop counterterms in the
gauge-less limit are thus given by

\begin{equation}
 \frac{\dMWsq}{M_W^2}=\frac{\Re\Self[W]{(1)}{0}}{M_W^2},\quad \frac{\dMZsq}{M_Z^2}=\frac{\Re\Self[Z]{(1)}{0}}{M_Z^2}.
 \label{Eq:dMVgless1L}
\end{equation}
Renormalization of the electric charge is not needed in the gauge-less limit.
Moreover, we do not need field renormalization because all the
field counterterms drop out in our calculation.

\section{\boldmath Corrections to the \texorpdfstring{$\rho$}{rho} parameter}
\label{Sec:rhoparameter}

The $\rho$ parameter 
\begin{equation}
\rho=\frac{G_{NC}}{G_{CC}}
\label{Eq:rhodefinition}
\end{equation}
was originally introduced \cite{Ross:1975fq}  for four-fermion processes at low momentum 
as the strength $G_{NC}$ of the effective neutral current coupling normalized by the charged current coupling $G_{CC}$.  
In the electroweak theory both classes of processes are mediated  by the exchange of a heavy gauge boson, 
the $Z$ boson for NC and the $W^\pm$ boson for CC processes.  
In the effective theory for low momentum transfer we can approximate 
the propagators by $1/M_V^2$ ($V=W,Z$). 
Therefore the effective couplings at the tree level are given by
\begin{align}
 \frac{G_{NC}}{\sqrt{2}}&=\frac{e^2}{8s_W^2c_W^2M_Z^2},\\
 \frac{G_{CC}}{\sqrt{2}}&=\frac{e^2}{8s_W^2M_W^2},
\end{align}
which results in
\begin{equation}
 \rho=\frac{M_W^2}{c_W^2 M_Z^2}=1.
\end{equation}
Including higher-order processes in the calculation of the effective couplings $G_{NC}$ and $G_{CC}$ leads to a deviation $\Delta \rho$ from unity,
\begin{equation}
\label{Eq:rhoparameter}
 \rho=\frac{1}{1-\Delta\rho} ,
\end{equation}
where
\begin{equation}
\label{Eq:delrhoperturbative}
\Delta\rho = \Delta\rho^{(1)} + \Delta\rho^{(2)} + \cdots  
\end{equation}
can be calculated in the loop-order expansion. Although conceptually defined at low-momentum scales,
the quantity $\Delta\rho$ represents an important ingredient for electroweak precision observables 
as the leading universal correction, with a substantial impact e.g.\ on the effective electroweak mixing angle and the $W$ mass.

Vertex and box-diagram corrections to charged and neutral current
processes do not contribute in the gauge-less limit 
and for vanishing masses of the external fermions, as well as  $\gamma$-$Z$ mixing in the neutral current interaction. 

Consequently, only corrections from the gauge-boson self-energies arise, of the form 
\begin{equation}
 \frac{\Self[V]{}{0}}{M_V^2}\quad(V=W,Z).
 \label{Eq:Selfratio}
\end{equation}
Due to a Ward identity in the gauge-less limit \cite{Barbieri:1992dq,Fleischer:1994cb} these quantities can be calculated also by the relations
\begin{align}
 \label{Eq:ZWardId}
 \frac{\Self[Z]{}{0}}{M_Z^2} & = -\Self[G^0]{\prime}{0}, \qquad
 \frac{\Self[W]{}{0}}{M_W^2} = -\Self[G^\pm]{\prime}{0},
\end{align}
where  the Goldstone self-energies are decomposed according to
\begin{equation}
 \Self[G]{}{p^2}=\Self[G]{}{0}+p^2 \Self[G]{\prime}{p^2}, \quad (G=G^0,G^\pm).
\end{equation}
We use this Ward identity as a test for our result. Moreover, the origin of a specific contribution in $\Delta\rho$ is not always directly visible in the calculation based on the gauge-boson self-energies due to the cancellation of the gauge couplings in the ratio (\ref{Eq:Selfratio}). In these cases, the couplings involved  can be identified with the help of the Ward identity.

\subsection{One-loop corrections in the SM and the THDM}

\begin{figure*}
 \begin{center}
  \includegraphics{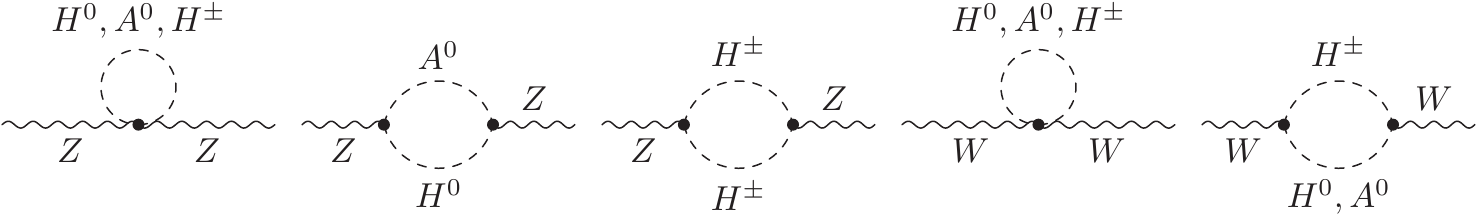}
  \caption{Non-standard contributions from the THDM scalars to the $Z$ and $W$ boson self-energies in the alignment limit at the one-loop level.}
  \label{Fig:ZW_Selfenergy_NS_1Loop}
 \end{center}
\end{figure*}

The calculation of the effective coupling strengths at the one-loop order in the gauge-less limit results in
\begin{equation}
\frac{G_{NC}}{\sqrt{2}}=\frac{e_0^2}{8s_{W,0}^2c_{W,0}^2M_{Z,0}^2}\left[1+\frac{\Self[Z]{(1)}{0}}{M_Z^2}\right]
\end{equation}
and
\begin{equation}
\frac{G_{CC}}{\sqrt{2}}=\frac{e_0^2}{8s_{W,0}^2M_{W,0}^2}\left[1+\frac{\Self[W]{(1)}{0}}{M_W^2}\right].
\end{equation}
The expansion of the bare parameters cancel in the $\rho$ parameter. Therefore, the correction to the $\rho$ parameter at one-loop order is given by
\begin{equation}
 \Delta \rho^{(1)}=\frac{\Self[Z]{(1)}{0}}{M_Z^2}-\frac{\Self[W]{(1)}{0}}{M_W^2}.
 \label{Eq:drho1Loopdef}
\end{equation}

In the THDM with the assumptions described in section~\ref{Sec:Outline} we can split the one-loop correction
\begin{equation}
 \Delta \rho^{(1)}=\Delrho[t]{1}+\Delrho[NS]{1}
 \label{Eq:drho1Loopsplit}
\end{equation}
into two independent parts originating from the top-Yukawa coupling and the scalar sector. The first part arises from the large mass splitting between the top and the bottom quark and is identical to the dominant part of the one-loop corrections to $\Delta\rho$ in the SM \cite{Veltman:1977kh,Chanowitz:1978mv,Chanowitz:1978uj}. When neglecting the mass of the bottom quark one obtains the one-loop result
\begin{equation}
\Delrho[t]{1}=\frac{3\alem}{16\pi M_W^2s_W^2}m_t^2.
\label{Eq:drhoSM1Loop}
\end{equation}

In the SM no contributions to $\Delta\rho^{(1)}$ arise from scalar
loops due to the custodial symmetry of the Higgs potential.  In a similar way there is no correction from the SM-like scalars $h^0$,
$G^0$ and $G^\pm$ in the THDM with the assumptions from section~\ref{Sec:Outline}, since the part $V_{I}$ of the potential in
\eqref{Eq:PotAlign} is custodial invariant. The  contributions from the
SM-like scalars to the gauge-boson self-energies yield the SM result 
in dimensional regularization with dimension $D$, expressed in terms
of $A_0$ in \ref{App:1LIntegrals},
\begin{equation}
\frac{\Self[V,\text{SM}]{(1)}{0}}{M_V^2}=\frac{\alem}{16 \pi  s_W^2 M_W^2}\frac{(D-4) }{ D }A_0\left(m_{h^0}^2\right),
\label{Eq:VVSM1Loop}
\end{equation}
for both $V=W,Z$.
They cancel in the difference for  $\Delta\rho^{(1)}$  in \eqref{Eq:drho1Loopdef}.

However, the extended scalar sector of the THDM gives additional
scalar contributions to $\Delta\rho$ 
\cite{Bertolini:1985ia,Hollik:1986gg,Hollik:1987fg,Denner:1991ie,Grimus:2007if,gunion:1990}. 
In the alignment limit the additional correction follows from the scalars $H^0$, $A^0$ and $H^\pm$. The gauge-boson self-energies from the diagrams in figure~\ref{Fig:ZW_Selfenergy_NS_1Loop}  give rise to the non-standard one-loop part   
\begin{align}
\Delrho[NS]{1}=&\frac{\alem}{16 \pi s_W^2 M_W^2 D }\Big\{4 m_{A^0}^2 B_0\left(0,m_{A^0}^2,m_{H^\pm}^2\right)\notag\\
  &+4 m_{H^0}^2 B_0\left(0,m_{H^0}^2,m_{H^\pm}^2\right)\notag\\
	&-4 m_{A^0}^2 B_0\left(0,m_{A^0}^2,m_{H^0}^2\right)\notag\\
	&+(8-2 D) A_0\left(m_{H^\pm}^2\right)-4A_0\left(m_{H^0}^2\right)\Big\}
 \label{Eq:drhoNS1L_arbD}
\end{align}
which simplifies to
\begin{align}
 \Delrho[NS]{1} \xrightarrow{D\rightarrow4}&\frac{\alem}{16 \pi  s_W^2 M_W^2}\Bigg\{  
\frac{m_{A^0}^2 m_{H^0}^2 }{m_{A^0}^2-m_{H^0}^2}\log \left(\frac{m_{A^0}^2}{m_{H^0}^2}\right)\notag\\
  &-\frac{m_{A^0}^2 m_{H^{\pm }}^2 }{m_{A^0}^2-m_{H^{\pm}}^2}\log \left(\frac{m_{A^0}^2}{m_{H^{\pm }}^2}\right)\notag\\
 &-\frac{m_{H^0}^2 m_{H^{\pm }}^2 }{m_{H^0}^2-m_{H^{\pm }}^2}\log \left(\frac{m_{H^0}^2}{m_{H^{\pm }}^2}\right)\notag\\
      &+m_{H^{\pm }}^2\Bigg\}.
\label{Eq:drhoNS1Loop}
\end{align}
in four dimensions.
It increases quadratically  with the mass difference between the charged and the neutral Higgs states, and it vanishes for 
\begin{equation}
\MHH=\MHp
\end{equation} 
or
\begin{equation}
\MAO=\MHp.
\end{equation}
The reason is that this correction contains only couplings between the Goldstone bosons and the non-standard scalars $H^0$, $A^0$ and $H^\pm$ which are determined by the part $V_{III}$ of the potential. As explained in section~\ref{Sec:CustSym}, the custodial symmetry in this part can be restored for equal charged and neutral Higgs masses.
Note that in the alignment case the entire non-standard one-loop
contribution to $\Delta\rho$ is exclusively given by
the expression~(\ref{Eq:drhoNS1Loop}), 
corresponding to the gauge-less limit.

\subsection{Higher-order corrections in the THDM}

As mentioned above, it is sufficient to 
keep only the corrections from the gauge-boson self-energies in the calculation of the effective neutral and charged current interaction of the four fermion processes. 
The two-loop results of the effective couplings are
\begin{align}
\frac{G_{NC}}{\sqrt{2}}
%=&\frac{e_0^2}{8s_{W,0}^2c_{W,0}^2M_{Z,0}^2}\left[1+\frac{\Self[Z]{(1)}{0}}{M_{Z,0}^2}+\left(\frac{\Self[Z]{(1)}{0}}{M_{Z}^2}\right)^2+\frac{\Self[Z]{(2)}{0}}{M_Z^2}\right]\\
	=&\frac{e_0^2}{8s_{W,0}^2c_{W,0}^2M_{Z,0}^2}\left[1+\frac{\Self[Z]{(1)}{0}}{M_{Z}^2}-\frac{\dMZsq}{M_Z^2}\frac{\Self[Z]{(1)}{0}}{M_{Z}^2}\right.\notag\\
	  &\left.+\left(\frac{\Self[Z]{(1)}{0}}{M_{Z}^2}\right)^2+\frac{\Self[Z]{(2)}{0}}{M_Z^2}\right]
\end{align}
and
\begin{align}
\frac{G_{CC}}{\sqrt{2}}
%=&\frac{e_0^2}{8s_{W,0}^2M_{W,0}^2}\left[1+\frac{\Self[W]{(1)}{0}}{M_{W,0}^2}+\left(\frac{\Self[W]{(1)}{0}}{M_{W}^2}\right)^2+\frac{\Self[W]{(2)}{0}}{M_W^2}\right]\\
	=&\frac{e_0^2}{8s_{W,0}^2M_{W,0}^2}\left[1+\frac{\Self[W]{(1)}{0}}{M_{W}^2}-\frac{\dMWsq}{M_W^2}\frac{\Self[W]{(1)}{0}}{M_{W}^2}\right.\notag\\
	  &\left.+\left(\frac{\Self[W]{(1)}{0}}{M_{W}^2}\right)^2+\frac{\Self[W]{(2)}{0}}{M_W^2}\right].
\end{align}
With the renormalization condition \eqref{Eq:dMVgless1L} for the gauge-boson mass
counterterms in the gauge-less limit the
products of one-loop corrections in the brackets cancel. 
The calculation of $\rho$ as defined by \eqref{Eq:rhodefinition}
then yields the deviation $\Delta\rho$ in \eqref{Eq:rhoparameter} as follows,

\begin{align}
\label{Eq:delroZW}
 \Delta \rho=& \left(\frac{\Self[Z]{(1)}{0}}{M_Z^2}-\frac{\Self[W]{(1)}{0}}{M_W^2}\right)\notag\\
  &-\frac{\Self[Z]{(1)}{0}}{M_Z^2}\left(\frac{\Self[Z]{(1)}{0}}{M_Z^2}-\frac{\Self[W]{(1)}{0}}{M_W^2}\right)\notag\\
    &+\left(\frac{\Self[Z]{(2)}{0}}{M_Z^2}-\frac{\Self[W]{(2)}{0}}{M_W^2}\right)\\
  =&\Delta\rho^{(1)}+\Delta\rho^{(2)},
\end{align}
where the two-loop part is given by 
\begin{equation}
 \Delta\rho^{(2)}\equiv-\frac{\Self[Z]{(1)}{0}}{M_Z^2}\Delrho{1}+\left(\frac{\Self[Z]{(2)}{0}}{M_Z^2}-\frac{\Self[W]{(2)}{0}}{M_W^2}\right) .
 \label{Eq:Deltarho2}
\end{equation}
$\Delta\rho^{(1)}$ summarizes the one-loop corrections as given by
\eqref{Eq:drho1Loopdef} and \eqref{Eq:drho1Loopsplit}. The self-energy of the $Z$
boson in the first term consists of all the corrections from the
top quark and the scalars as internal particles. Note that it contains
also the part from the SM-like scalars in \eqref{Eq:VVSM1Loop}, which
cancel in $\Delrho{1}$. The second part of \eqref{Eq:Deltarho2} follows
from the two-loop corrections to the gauge-boson self-energies. In
addition to the part from the genuine two-loop diagrams (labeled as $\delrho{2Loop}$) 
it also includes one-loop diagrams with counterterm insertions for the subloop
renormalization (labeled as $\Delrho{CT}$).

With the assumptions from section~\ref{Sec:Outline} we have two sources for the two-loop contribution $\Delrho{2}$: the top-Yukawa interaction and the scalar self-interaction. 
Due to the alignment limit we can subdivide the top-Yukawa corrections into two parts. The first one is identical to the two-loop top-Yukawa contribution in the SM and is discussed in section~\ref{Subsec:drhoSMt}. The second one originates from the coupling between the top quark and the non-standard scalars $H^0$, $A^0$ and $H^\pm$ and is described in more detail in section~\ref{Subsec:drhoNSt}.
A similar separation can be made for the additional corrections to the $\rho$ parameter from the scalar self-interaction. The part $V_I$ of the potential (see \eqref{Eq:V_I}), which describes only the interaction between $h^0$ and the Goldstone bosons $G^0$, $G^\pm$, is invariant under the custodial symmetry and the corresponding contributions to the vector-boson self-energies in $\Delta\rho$ cancel each other. The remaining part of the potential gives rise to two finite subsets in $\Delrho{2}$. One follows from the interaction between the SM-like scalars $h^0$, $G^0$, $G^\pm$ and the non-standard scalars $H^0$, $A^0$, $H^\pm$ and is discussed in section~\ref{Subsec:drhoMix}. The other one contains only the non-standard scalars $H^0$, $A^0$ and $H^\pm$ as internal particles in the gauge-boson self-energies and is described in section~\ref{Subsec:drhoNSHiggs}.

With this categorization we subdivide the contribution from the
genuine two-loop diagrams (without subloop renormalization) to the
vector-boson self-energies into different parts, according to their origin,
\begin{multline}
 \delrho{2Loop}=\delrho[t,SM]{2Loop}+\delrho[t,NS]{2Loop}\\
 +\delrho[H,NS]{2Loop}+\delrho[H,Mix]{2Loop}
 \label{Eq:dr2Loopsplit}
\end{multline}
which are classified by the participating couplings:
\begin{itemize}
 \item $\delrho[t,SM]{2Loop}$ originates from the coupling between the top quark and the SM-like scalars $h^0$, $G^0$ and $G^\pm$ (see section~\ref{Subsec:drhoSMt});
 \item $\delrho[t,NS]{2Loop}$ is the part which follows from the top-Yukawa interaction of the non-standard scalars $H^0$, $A^0$ and $H^\pm$ (see section~\ref{Subsec:drhoNSt});
 \item $\delrho[H,NS]{2Loop}$ contains the scalar self-coupling between the non-standard scalars (see section~\ref{Subsec:drhoNSHiggs});
 \item $\delrho[H,Mix]{2Loop}$ follows from the interaction between the SM-like scalars and the non-standard scalars (see section~\ref{Subsec:drhoMix}).  
\end{itemize}

\begin{figure}[ht]
 \begin{center}
  \includegraphics[width=0.45\textwidth]{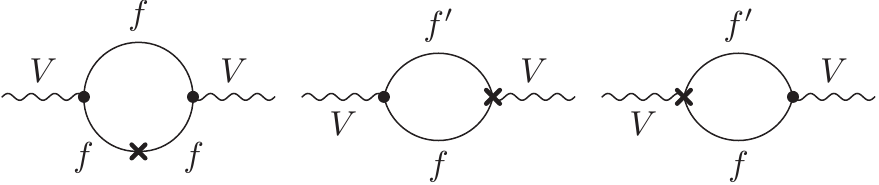}
  \caption{Generic diagrams for the gauge-boson self-energies containing quarks with counterterm insertions. $V=\{W,Z\}$; $f,f^\prime=\{t,b\}$.}
  \label{Fig:VV_t_CTDiagramme}
 \end{center}
\end{figure}

\begin{figure*}[ht]
 \begin{center}
  \includegraphics{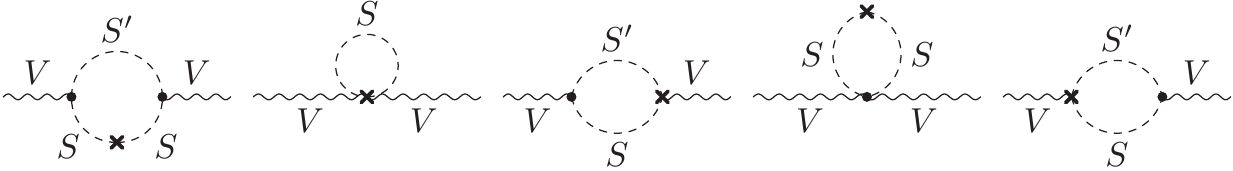}
  \caption{Generic diagrams for the gauge-boson self-energies $V=\{W,Z\}$ containing  scalars with counterterm insertions. 
   The  contribution from the diagrams can be divided into two parts: one part with only the SM-like scalars ($S,S^\prime=\{h^0,G^0,G^\pm\}$) 
   and one part with only the non-standard scalars ($S,S^\prime=\{H^0, A^0, H^\pm\}$).}
  \label{Fig:VV_NS_CTDiagramme}
 \end{center}
\end{figure*}

For one-loop subrenormalization we need the diagrams shown in figure~\ref{Fig:VV_t_CTDiagramme} for the
self-energies with the top quarks and in figure~\ref{Fig:VV_NS_CTDiagramme} for the scalar contribution. In the gauge-less limit only two types of
renormalization constants survive: the counterterm $\dSWsq$ from the  counterterm insertions in the vertices, and the mass counterterms in
the propagators of the internal particles. All field counterterms of the internal particles drop out in the calculation, 
and all other counterterms are zero in the gauge-less limit.

From the diagrams of figure~\ref{Fig:VV_t_CTDiagramme} we obtain the part of
the subloop renormalization from the top quark. The renormalization
of the weak mixing angle is contained in the vertex counterterms
(see \ref{App:Feynmanrules}) and yields the term  
\begin{equation}
\frac{s_W^2}{c_W^2}\frac{\dSWsq}{s_W^2}\frac{\Self[Z,\text{t}]{(1)}{0}}{M_Z^2}-\frac{\dSWsq}{s_W^2}\Delrho[t]{1} .
\label{Eq:VertexCTtop}
\end{equation}
From the diagrams with counterterms  in the propagators in figure~\ref{Fig:VV_t_CTDiagramme} we obtain the term
\begin{equation}
 \delrho[t]{CT}=-\frac{3 \alem  (D-4) (D-2)^2  A_0\left(m_t^2\right)}{16 \pi  D  M_W^2 s_W^2}\frac{\delta m_t}{m_t}.
 \label{Eq:drhotCT}
\end{equation}
Due to the alignment limit we can split the result of the top mass counterterm into a SM-like and a non-standard part. We use this for the seperation
\begin{equation}
\delrho[t]{CT}=\delrho[t,SM]{CT}+\delrho[t,NS]{CT} \, ,
\end{equation}
where the two parts are defined as follows:
\begin{itemize}
  \item the part $\delrho[t,SM]{CT}$ contains the correction to the top-mass counterterm from the SM-like scalars $h^0$, $G^0$, $G^\pm$ as shown in the self-energy diagrams in figure~\ref{Fig:topSMCT};
  \item the second part $\delrho[t,NS]{CT}$ contains the part of $\delta m_t$ which comes from the top quark self-energy corrections from the non-standard scalars as depicted in figure~\ref{Fig:topNSCT}.
\end{itemize}

For the subloop renormalization diagrams in figure~\ref{Fig:VV_NS_CTDiagramme} with the SM-like scalars $h^0$, $G^0$ and $G^\pm$ we find that the mass counterterms drop out in the difference of the $W$ and $Z$ self-energy, due to custodial symmetry.  From the vertex counterterms we obtain the contribution
\begin{equation}
\frac{s_W^2}{c_W^2}\frac{\dSWsq}{s_W^2}\frac{\Self[Z,\text{SM}]{(1)}{0}}{M_Z^2}
\label{Eq:VertexCTSM}
\end{equation}
with the one-loop self-energy from \eqref{Eq:VVSM1Loop}.

The diagrams in figure~\ref{Fig:VV_NS_CTDiagramme} with the possible
insertions of the non-standard scalars for $S$ and $S^\prime$ give the
last part of the subloop renormalization. With the Feynman rules of
\ref{App:Feynmanrules} the counterterms in the vertices yield the contribution
\begin{equation}
 \frac{\Self[Z,\text{NS}]{(1)}{0}}{M_Z^2}\frac{s_W^2}{c_W^2}\frac{\dSWsq}{s_W^2}-\frac{\dSWsq}{s_W^2}\Delrho[NS]{1},
 \label{Eq:VertexCTNS}
\end{equation}
where the $Z$ self-energy in the first term contains just the contribution of the non-standard scalars.

The correction from the mass counterterms $\dMHHsq$, $\dMAOsq$ and $\dMHpsq$ in the diagrams in figure~\ref{Fig:VV_NS_CTDiagramme} is denoted by $\delrho[H]{CT}$. It is identical to 
\begin{equation}
\begin{aligned}
 \delrho[H]{CT}=\Big(&\delta m_{A^0}^2\frac{\partial}{\partial m_{A^0}^2}+\delta m_{H^0}^2\frac{\partial}{\partial m_{H^0}^2}\\
  &+\delta m_{H^\pm}^2\frac{\partial}{\partial m_{H^\pm}}\Big)\Delrho[1]{NS},
  \label{Eq:drhoNSCT}
\end{aligned}
 \end{equation}
 with the one-loop contribution from \eqref{Eq:drhoNS1L_arbD}.
By splitting up  the mass counterterms we will classify three different parts
\begin{equation}
\delrho[H]{CT}=\delrho[H,t]{CT}+\delrho[H,NS]{CT}+\delrho[H,Mix]{CT} \, ,
\label{Eq:drhoHCT}
\end{equation}
which are defined as follows:
\begin{itemize}
	\item $\delrho[H,t]{CT}$ contains the non-standard scalar mass
          counterterms originating from the top-Yukawa coupling. The corresponding diagrams are shown in figure~\ref{Fig:SStopCT}.
	\item $\delrho[H,NS]{CT}$ labels the part which contains only
          non-standard scalars in the calculation of $\dMHHsq$, $\dMAOsq$ and $\dMHpsq$. 
         The diagrams are displayed in figure~\ref{Fig:SSNSCTs}. 
	\item $\delrho[H,Mix]{CT}$ incorporates the contribution to the mass counterterms of $H^0$, $A^0$ and $H^\pm$ which originates from the couplings of the non-standard scalars to the SM-like scalars. The corresponding self-energy diagrams are presented in figure~\ref{Fig:SSMixCTs}.
\end{itemize}
When we combine the various parts from the subloop renormalization,
their overall contribution to $\Delrho{2}$ can be written as follows:
\begin{equation}
\begin{aligned}
 \Delrho{CT}=&\frac{s_W^2}{c_W^2}\frac{\dSWsq}{s_W^2}\frac{\Self[Z]{(1)}{0}}{M_Z^2}\\
  &-\frac{\dSWsq}{s_W^2}\left(\Delrho[t]{1}+\Delrho[NS]{1}\right)+\delrho{CT}.
 \end{aligned}
 \label{Eq:DeltarhoCT}
\end{equation}
The first term incorporates all  parts from \eqref{Eq:VertexCTtop}, \eqref{Eq:VertexCTSM}  and \eqref{Eq:VertexCTNS} 
involving a single $Z$-boson self-energy; the remaining terms from the renormalization of $s_W$ in
\eqref{Eq:VertexCTtop} and \eqref{Eq:VertexCTNS} are kept separately in the second term. The last term
\begin{equation}
 \delrho{CT}=\delrho[t]{CT}+\delrho[H]{CT}
 \label{Eq:drhoCT} 
\end{equation}
collects the various parts resulting from the mass counterterms of the internal particles.

The two-loop correction to the $\rho$ parameter in \eqref{Eq:Deltarho2} can be further simplified, 
since the counterterm of the weak mixing angle reduces to
\begin{equation}
\frac{\dSWsq}{s_W^2}=\frac{c_W^2}{s_W^2}\left(\frac{\Self[Z]{(1)}{0}}{M_Z^2}-\frac{\Self[W]{(1)}{0}}{M_W^2}\right)=\frac{c_W^2}{s_W^2}\Delrho{1}{}. 
\label{Eq:dSWgless}
\end{equation}
in the gauge-less limit (see \eqref{Eq:dMVgless1L}). 
Combined  with \eqref{Eq:DeltarhoCT} the first term in \eqref{Eq:Deltarho2} is canceled and we obtain
\begin{equation}
 \Delta\rho^{(2)}=-\frac{c_W^2}{s_W^2}\left(\Delrho{1}\right)^2+\delrho{2} .
\label{Eq:withreduciblepart}
\end{equation}
 In this notation, the genuine two-loop part 
\begin{equation}
\label{Eq:delro2genuine}
\delrho{2}=\delrho{CT}+\delrho{2Loop}
\end{equation}
contains $\delrho{CT}$ resulting exclusively from the insertions of the mass counterterms, 
and the contribution $\delrho{2Loop}$ from the pure two-loop diagrams  
for the $Z,W$ self-energies (without subloop renormalization) in \eqref{Eq:Deltarho2}.

The appearance of the reducible term $\left(\Delrho{1}\right)^2$ in $\Delrho{2}$ is a consequence of the parameterisation of $v^2$ by
\begin{equation}
 \frac{1}{v^2}=\frac{e^2}{4 s_W^2 M_W^2}
\end{equation}
together with the on-shell renormalization of $s_W$. A different parameterisation in terms of the Fermi constant~$G_F$ 
can be introduced  with the help of the relation 
\begin{equation}
 \sqrt{2} G_F=\frac{e^2}{4 M_W^2 s_W^2}\left(1+\Delta r\right) ,
\end{equation}
 where the quantity $\Delta r$ describes the higher-order corrections. In the gauge-less limit the one-loop contribution is given by
\begin{equation}
 \Delta r =-\frac{\dSWsq}{s_W^2}.
\end{equation}
Consequently, the reparameterisation of the one-loop result $\Delrho{1}$ in terms of $G_F$ induces a two-loop shift 
originating from $\Delta r$, which effectively cancels the reducible
term in $\Delrho{2}$ in~\eqref{Eq:withreduciblepart}.
Hence, in the $G_F$ expansion, the two-loop contribution in $\Delta\rho$ 
is identified as the irreducible two-loop part $\delta\rho^{(2)}$ in~\eqref{Eq:delro2genuine}.
In this way, the same pattern for $\rho$  is found
as in the SM \cite{Consoli:1989fg}.

The structure of the irreducible quantity $\delta\rho^{(2)}$ in \eqref{Eq:delro2genuine} 
with $\delta\rho^{({\rm 2Loop})}$ defined in \eqref{Eq:dr2Loopsplit}
allows us to divide it into four finite subsets of different origins,
\begin{equation}
\label{Eq:aufteilung}
 \delta\rho^{(2)} = \delta\rho^{(2)}_{\rm t, SM} + \delta\rho^{(2)}_{\rm t, NS} 
 + \delta\rho^{(2)}_{\rm H,NS} + \delta\rho^{(2)}_{\rm H, Mix} \, ,
\end{equation}
which we describe now in more detail.

\begin{figure*}[ht]
 \begin{center}
  \includegraphics{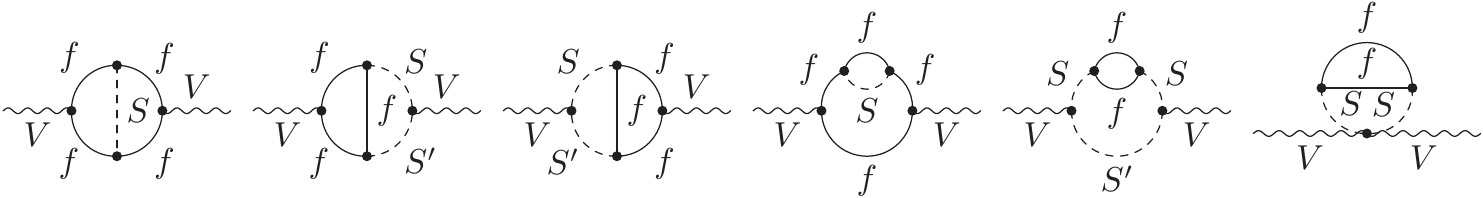}
  \caption{Generic two-loop diagrams for the top-Yukawa corrections to the vector-boson self-energies with $V=\{W,Z\}$ and $f=\{t,b\}$. The standard contribution $\delrho[t,SM]{2Loop}$ follows from $S,S^\prime=\{h^0, G^0, G^\pm\}$. The non-standard contribution $\delrho[t,NS]{2Loop}$ is obtained by all possible insertions of $S,S^\prime=\{H^0, A^0, H^\pm\}$.}
  \label{Fig:NStop2Loop}
  \end{center}
\end{figure*}

\subsubsection{Standard model corrections from the top-Yukawa coupling}
\label{Subsec:drhoSMt}

\begin{figure}[ht]
 \begin{center}
  \includegraphics{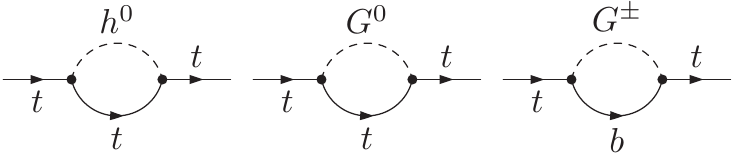}
   \caption{One-loop diagrams for the standard contribution to the top-mass counterterm $\delta m_t$.}
  \label{Fig:topSMCT}
 \end{center}
\end{figure} 

The first contribution under investigation are the two-loop corrections from the top-Yukawa coupling. In the alignment limit this corrections can be split into two independent subsets. From the coupling of the top quark to the SM-like scalars $h^0$, $G^0$ and $G^\pm$ we obtain the finite correction
\begin{equation}
 \delrho[t,SM]{2}=\delrho[t,SM]{CT}+\delrho[t,SM]{2Loop}.
\end{equation}
$\delrho[t,SM]{2Loop}$ are the pure two-loop contributions, which are depicted by the generic diagrams in figure~\ref{Fig:NStop2Loop} for $S,S^\prime={h^0,G^0, G^\pm}$. Its divergences are cancelled by the part $\delrho[t,SM]{CT}$ which is the part of \eqref{Eq:drhotCT} with the top-mass counterterm calculated from the diagrams in figure~\ref{Fig:topSMCT}. $\delrho[t,SM]{2}$ is identical to the already known SM contribution from the top-Yukawa interaction. First the result was calculated in the approximation $M_{H}=0$ \cite{vanderBij:1986hy} and as an expansion for large values of $M_{H}$ \cite{vanderBij:1983bw}. Later the full result for arbitrary Higgs masses was obtained \cite{Barbieri:1992dq,Fleischer:1993ub,Fleischer:1994cb}. We checked that our calculation leads to the same result.

\subsubsection{Non-standard corrections from the top-Yukawa coupling}
\label{Subsec:drhoNSt}

\begin{figure}[t]
 \begin{center}
  \includegraphics{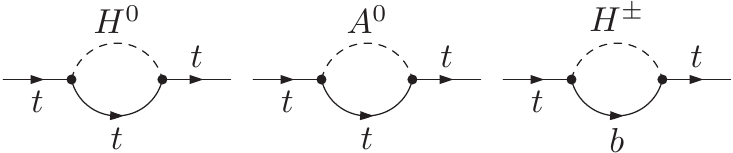}
   \caption{One-loop diagrams for the non-standard contribution to the top-mass counterterm $\delta m_t$.}
  \label{Fig:topNSCT}
 \end{center}
\end{figure} 
 \begin{figure}[t]
 \begin{center} 
  \includegraphics{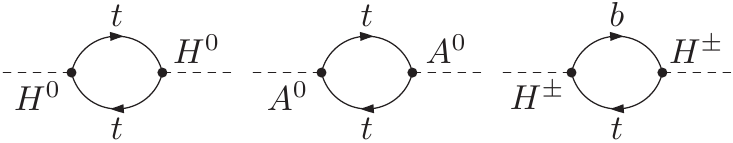}
  \caption{One-loop diagrams for the top-Yukawa contribution to the non-standard scalar mass counterterms.}
  \label{Fig:SStopCT}
 \end{center}
\end{figure}

\noindent
More interesting is the additional contribution due to the coupling of the top quark to the non-standard scalars $H^0$, $A^0$ and $H^\pm$, which is given by
\begin{equation}
 \delrho[t,NS]{2}=\delrho[t,NS]{CT}+\delrho[H,t]{CT}+\delrho[t,NS]{2Loop}.
\end{equation}
$\delrho[t,NS]{2Loop}$ denotes the pure two-loop part, represented by
the generic diagrams shown in figure~\ref{Fig:NStop2Loop} with
$S,S^\prime=\{H^0,A^0,H^\pm\}$. The result does not only consist of terms
of $\order{\alpha_t^2}$, which originate only from the top-Yukawa
interaction, but also of contributions of $\order{\alpha_t \lambda_i}$ which contain the scalar self-couplings in addition to the top-Yukawa coupling. The divergences from the $\order{\alpha_t^2}$ part are canceled by $\delrho[t,NS]{CT}$ which originates from the subloop renormalization diagrams of figure~\ref{Fig:VV_t_CTDiagramme} with the top-mass counterterm calculated from the diagrams in figure~\ref{Fig:topNSCT}. The divergences of $\order{\alpha_t\lambda_i}$ are cancelled by $\delrho[H,t]{CT}$ with the mass counterterms calculated from the diagrams in figure~\ref{Fig:SStopCT}. 
In the calculation by means of the gauge-boson self-energies the
separation between the $\order{\alpha_t^2}$ and the $\order{\alpha_t  \lambda_i}$ contributions is obscured. 
Using the Ward identity in \eqref{Eq:ZWardId} can help to disentangle the
two different finite contributions of $\order{\alpha_t^2}$ and $\order{\alpha_t\lambda_i}$.

\subsubsection{Scalar corrections from the interaction of the non-standard scalars}
\label{Subsec:drhoNSHiggs}

\begin{figure}[ht]
\begin{center}
 \includegraphics{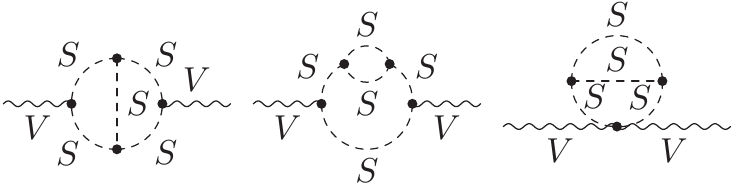}
\caption{Generic two-loop diagrams for the vector-boson self-energies from the interaction from the non-standard scalars. $V=\{W,Z\}$; $S=\{H^0, A^0, H^\pm\}$.}
\label{Fig:NSHiggs2Loop}
\end{center}
\end{figure}

\begin{figure}[ht]
 \begin{center}
  \includegraphics{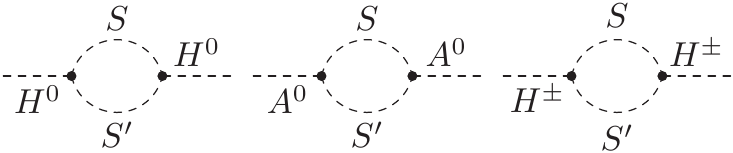}
  \caption{One-loop diagrams for the non-standard scalar mass counterterms from the interaction between the non-standard scalars. 
    For the $H^0$ self-energy: $S=S^\prime=H^0,A^0,H^\pm$. For the $A^0$ self-energy: $S=A^0$ and $S^\prime=H^0$. For the $H^\pm$ self-energy: $S=H^\pm$ and $S^\prime=H^0$.} 
  \label{Fig:SSNSCTs}
 \end{center}
\end{figure}

The interaction between the non-standard scalars gives another finite subset. When inspecting this contribution we found that all the corrections from a coupling between four non-standard scalars are cancelled. The two-loop diagrams which contain such a coupling can be written as a product of two scalar one-loop integrals. The mass counterterms in the subloop renormalization lead to the same product from the corrections to the scalar self-energies, but with an opposite sign. Consequently the two terms cancel each other.

The remaining contribution
\begin{equation}
\delrho[H,NS]{2}=\delrho[H,NS]{CT}+\delrho[H,NS]{2Loop} 
\end{equation}
comes from all the diagrams which include a triple scalar coupling between $H^0$, $A^0$ and $H^\pm$. $\delrho[H,NS]{2Loop}$ is the result for the vector-boson self-energies of the generic two-loop diagrams in figure~\ref{Fig:NSHiggs2Loop}. For the subloop renormalization we need the corrections from the triple non-standard scalar coupling to the scalar self-energies, as shown in figure~\ref{Fig:SSNSCTs}. Inserting the corresponding mass counterterms into \eqref{Eq:drhoNSCT} leads to the result of $\delrho[H,NS]{CT}$.

\subsubsection{Scalar corrections from the interaction of the non-standard scalars with the SM scalars}\label{Subsec:drhoMix}

\begin{figure*}[ht]
\begin{center}
 \includegraphics{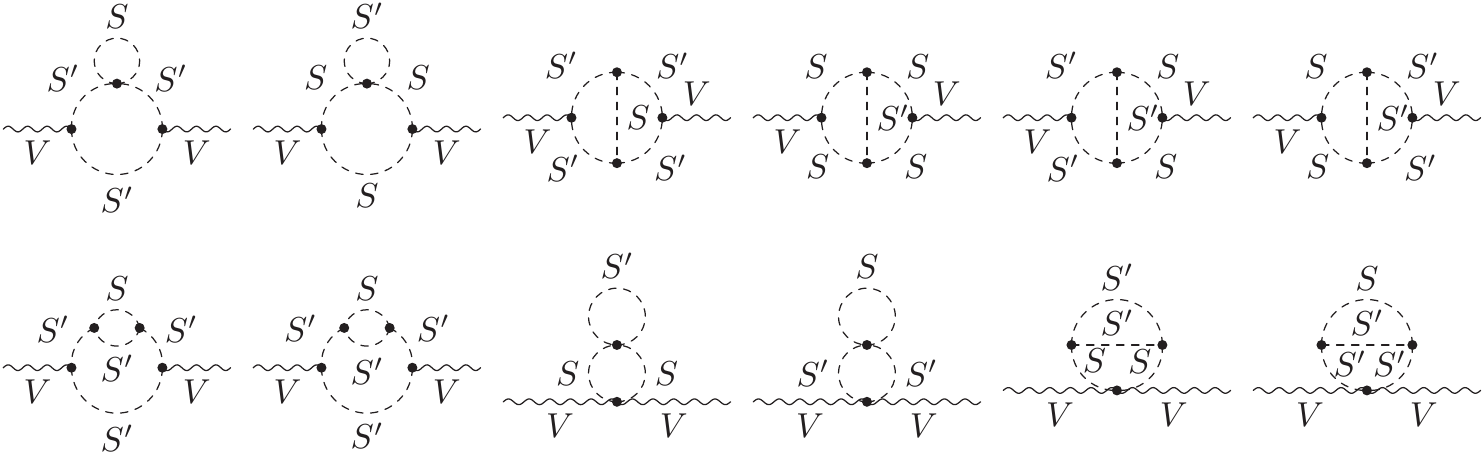}
 \caption{Generic two-loop diagrams from the interaction between the SM-like scalars $S=h^0$, $G^0$, $G^\pm$ and the non-standard scalars $S^\prime=H^0$, $A^0$, $H^\pm$. $V=\{W,Z\}$}
 \label{Fig:Mix2Loop}
 \end{center}
\end{figure*}

\begin{figure*}[ht]
 \begin{center}
  \includegraphics{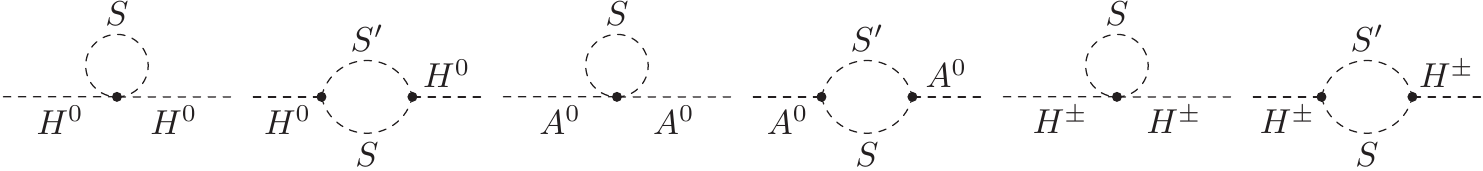}
\caption{One-loop diagrams for the non-standard scalar mass counterterms 
from the interaction between the SM-like scalars $S=h^0, G^0, G^\pm$ and the non-standard scalars $S^\prime=H^0,A^0,H^\pm$.}
\label{Fig:SSMixCTs}
\end{center}
\end{figure*}

As already mentioned another finite subset of two-loop corrections to the $\rho$ parameter comes from the interaction between the scalars $h^0$, $G^0$, $G^\pm$ with the non-standard scalars $H^0$, $A^0$, $H^\pm$. This interaction follows only from the part $V_{III}$ of the potential (see \eqref{Eq:V_III}) which is custodial-symmetry breaking. We denote the resulting contribution by
\begin{equation}
 \delrho[H,Mix]{2}=\delrho[H,Mix]{CT}+\delrho[H,Mix]{2Loop},
\end{equation}
where $\delrho[H,Mix]{2Loop}$ is the part from the two-loop diagrams shown in figure~\ref{Fig:Mix2Loop}. 
The divergences are canceled by $\delrho[H,Mix]{CT}$ from \eqref{Eq:drhoHCT}, which is obtained by calculating the mass counterterms in \eqref{Eq:drhoNSCT} from the diagrams in figure~\ref{Fig:SSMixCTs}.

\subsection{The Inert-Higgs-Doublet model}
\label{Sec:IHDM}

We now discuss our result in the context of a special version of the THDM,  the Inert-Higgs-Doublet Model (IHDM).
Originally proposed in~\cite{Deshpande:1977rw}, it has received topical attention in the light of 
neutrino mass phenomenology  and dark-matter searches (see e.g.~\cite{Arhrib:2013ela}
for a recent comprehensive analysis and more references).
In the IHDM the SM scalar sector is extended by a second complex doublet with the special feature that the Lagrangian has an exact $Z_2$ symmetry under which all the SM particles are even while the second doublet is odd. This $Z_2$ symmetry has several interesting consequences. The requirement that it stays unbroken forbids a vacuum expectation value of the second doublet. Therefore the doublets of the IHDM are
\begin{eqnarray}
 H_1=&
 \begin{pmatrix}
  G^\pm \\
  \frac{1}{\sqrt{2}}\left(v+h^0+iG^0\right)
 \end{pmatrix},\\
 H_2=&
 \begin{pmatrix}
  H^\pm\\
  \frac{1}{\sqrt{2}}\left(H^0+iA^0\right)
 \end{pmatrix}.
\end{eqnarray}
The doublet $H_1$ is identical to the scalar doublet in the SM. It consists of the physical SM-like Higgs boson $h^0$ and the Goldstone bosons $G^0$ and $G^\pm$. The second doublet $H_2$ transforms under the $Z_2$ as $H_2\rightarrow -H_2$ and contains the $CP$-even scalar $H^0$, the $CP$-odd scalar $A^0$ and the charged scalars $H^\pm$.  All the terms in the Lagrangian in which the SM particles couple to a single scalar of $H_2$ are forbidden by the $Z_2$ symmetry and the lightest of the scalars from $H_2$ is stable. If this is one of the neutral states $H^0$ or $A^0$ the IHDM provides a dark matter candidate.

\newcommand{\VIHDM}[1][{}]{V^\text{IHDM}_{#1}}
The most general scalar potential which is renormalizable, gauge invariant and respects the $Z_2$ symmetry is given by (see for example \cite{Arhrib:2015hoa})   
\begin{align}
\VIHDM=&\mu _1^2 \left(H_1^{\dagger }H_1\right)+\mu
   _2^2 \left(H_2^{\dagger } H_2\right)+\Lambda _1 \left(H_1^{\dagger } H_1\right){}^2 \notag\\
    &+\Lambda _2\left(H_2^{\dagger } H_2\right){}^2 +\Lambda _3 \left(H_2^{\dagger } H_2\right) \left(H_1^{\dagger } H_1\right)\notag\\
    &+\Lambda _4 \left(H_1^{\dagger } H_2\right) \left(H_2^{\dagger } H_1\right)\notag\\
    &+\frac{1}{2} \Lambda _5
   \left(\left(H_1^{\dagger } H_2\right){}^2+\left(H_2^{\dagger } H_1\right){}^2\right) .
\end{align} 
To avoid $CP$-violation all the parameters in the potential are chosen to be real. The minimization condition can be used to eliminate one of the parameters from the potential. From the remaining six parameters four can be expressed by the scalar masses (see \cite{Arhrib:2015hoa} for the explicit relations between the masses and the potential parameters). If we choose $\mu_2^2$ and $\Lambda_2$ for the remaining two parameters we can express the potential by   
\begin{align}
   \VIHDM=&\VIHDM[I]+\VIHDM[II]+\VIHDM[III] , \\[0.2cm]
   \VIHDM[I]=&\frac{m_{h^0}^2 }{2 v^2}\left(H_1^{\dagger } H_1\right){}^2-\frac{1}{2} m_{h^0}^2 \left(H_1^{\dagger } H_1\right) , \\
   \VIHDM[II]=&\mu _2^2 \left(H_2^{\dagger } H_2\right)+\Lambda _2 \left(H_2^{\dagger } H_2\right){}^2 , \\
   \VIHDM[III]=&\frac{ \left(m_{A^0}^2-2 m_{H^\pm}^2+m_{H^0}^2\right)}{v^2}\left(H_1^{\dagger } H_2\right) \left(H_2^{\dagger } H_1\right)\notag\\
    &+\frac{\left(m_{H^0}^2-m_{A^0}^2\right)}{2 v^2}\left[\left(H_1^{\dagger } H_2\right){}^2+\left(H_2^{\dagger } H_1\right){}^2\right]\notag\\
    &+\frac{2  \left(m_{H^\pm}^2-\mu _2^2\right)}{v^2}\left(H_1^{\dagger } H_1\right) \left(H_2^{\dagger } H_2\right).
\end{align}

We see that the SM-like doublet $\PhiSM$ in the aligned THDM is identical to the doublet $H_1$ in the IHDM. The non-standard doublet $\PhiNS$ in \eqref{Eq:NSdoublet} differs from the doublet $H_2$ of the IHDM by the overall sign in front of the $CP$-even scalar $H^0$. However, our result is independent on this overall sign, since the $CP$-even scalar $H^0$ appears only as an internal particle in the calculated self-energies. Therefore we can identify the doublet $H_2$ with the doublet $\PhiNS$. By using this identifications we can relate the potential between the IHDM and the more general THDM in the alignment limit in order to interpret our results in the context of the IHDM: 
\begin{itemize}
\item There is no non-standard correction to $\Delta\rho$ from the top-Yukawa interaction, since the interaction of the fermions with the non-standard scalars is forbidden by the $Z_2$ symmetry.
\item The part $\VIHDM[I]$ has the same structure as the scalar potential of the SM and will not lead to contributions to the $\rho$ parameter since it is invariant under the custodial symmetry (see section~\ref{Sec:CustSym}). 
\item In the IHDM all the quartic couplings between four non-standard scalars are proportional to $\Lambda_2$. However, in our calculation in the aligned THDM we found that all the contributions to $\Delta\rho$ from couplings between four non-standard scalars vanish (see section~\ref{Subsec:drhoNSHiggs}). The responsible arguments can also be transferred to the IHDM.  
\item When we identify $H_1$ with $\PhiSM$ and $H_2$ with $\PhiNS$ we see that the part $V_{III}$ of the potential in the aligned THDM can be obtained by the replacement 
\begin{equation}
\mu_2^2 = \frac{1}{2}\lambda_5 v^2- \frac{ m_{h^0}^2}{2}
\label{Eq:mu2relation}
\end{equation}
in $\VIHDM[III]$. Consequently for the calculation of the $\rho$ parameter in the IHDM we get corrections corresponding to $\Delrho[NS]{1}$ and $\delrho[H,Mix]{2}$. The one-loop correction $\Delrho[NS]{1}$ is identical in the IHDM since it is independent of $\lambda_5$. The two-loop part $\delrho[H,Mix]{2}$ can be written in terms of the IHDM parameter $\mu_2^2$ by using \eqref{Eq:mu2relation}.  
\item As mentioned in section~\ref{Subsec:drhoNSHiggs}, the correction $\delrho[H,NS]{2}$ contains the interaction between three of the non-standard scalars $H^0$, $A^0$ and $H^\pm$ which follows from the part $V_{IV}$ of the potential in \eqref{Eq:PotAlign}. In the IHDM couplings between three non-standard scalars are forbidden because of the exact $Z_2$ symmetry. As a consequence, corrections to the $\rho$ parameter which would correspond to $\delrho[H,NS]{2}$ are absent in the IHDM.
\end{itemize}

\section{Numerical results}
\label{Sec:Results}

In this part we present the numerical results of the two-loop corrections to the $\rho$ parameter. 
We study the dependence on the various parameters  of the aligned THDM 
and compare the non-standard two-loop contributions with the one-loop result
which is part of existing calculations of electroweak precision observables so far.
In this way the parameter regions emerge where the one-loop calculations
are insufficient and bounds on parameters derived from experimental precision data
will be significantly changed when the two-loop terms are taken into account.  
%
%In the on-shell scheme the vacuum expectation is parametrized by
%
%\begin{equation}
% \frac{1}{v^2}=\frac{e^2}{4 s_W^2 M_W^2}=\frac{\alem \pi}{M_W^2 s_W^2}.
%\end{equation}
The values for the SM input parameters are \cite{Olive:2016xmw} 
\begin{eqnarray}
 M_W  =&80.385 \text{ GeV}, \\
 M_Z =&91.1876 \text{ GeV}, \\
 m_t =&173.21 \text{ GeV}.
\end{eqnarray}
For the mass of the SM-like Higgs state $h^0$ we take over the value $m_{h^0} = 125\, \text{GeV}$.

The effect of non-standard corrections to electroweak observables is often parametrized in terms of the parameter 
set $S$, $T$,  $U$, originally defined in \cite{Peskin:1990zt,Peskin:1991sw}. 
Following the conventions of~\cite{Olive:2016xmw},  the quantity $T$ is related   
to the correction $\Delta\rho$ via
\begin{equation}
 \Delta\rho=\hat{\alpha}\left(M_Z\right)T
\end{equation}
with the running electromagnetic fine structure constant\cite{Olive:2016xmw}
\begin{equation}
 \hat{\alpha}\left(M_Z\right)^{-1}=127.950\pm0.017 \, .
\end{equation}
%is the fine-structure constant in the $\overline{\text{MS}}$ scheme evaluated at $M_Z$. 
The  current value of $T$ ~\cite{Olive:2016xmw},
 determined  from experimental data,
\begin{equation}
\label{Eq:Texp}
 T=0.08\pm0.12 \, ,
\end{equation}
can be translated into bounds for $\Delta\rho$ according to
\begin{equation}
 -0.000313\leq \Delta\rho \leq 0.00156,
\end{equation}
which can be used for a quick estimate of the effect of the 
higher-order contributions to $\Delta\rho$ in view of current
experimental constraints.

\subsection{Results for the top-Yukawa contribution}
\label{Sec:NStResults}

\begin{figure}[ht]
\begin{center}
 \includegraphics[width=\linewidth]{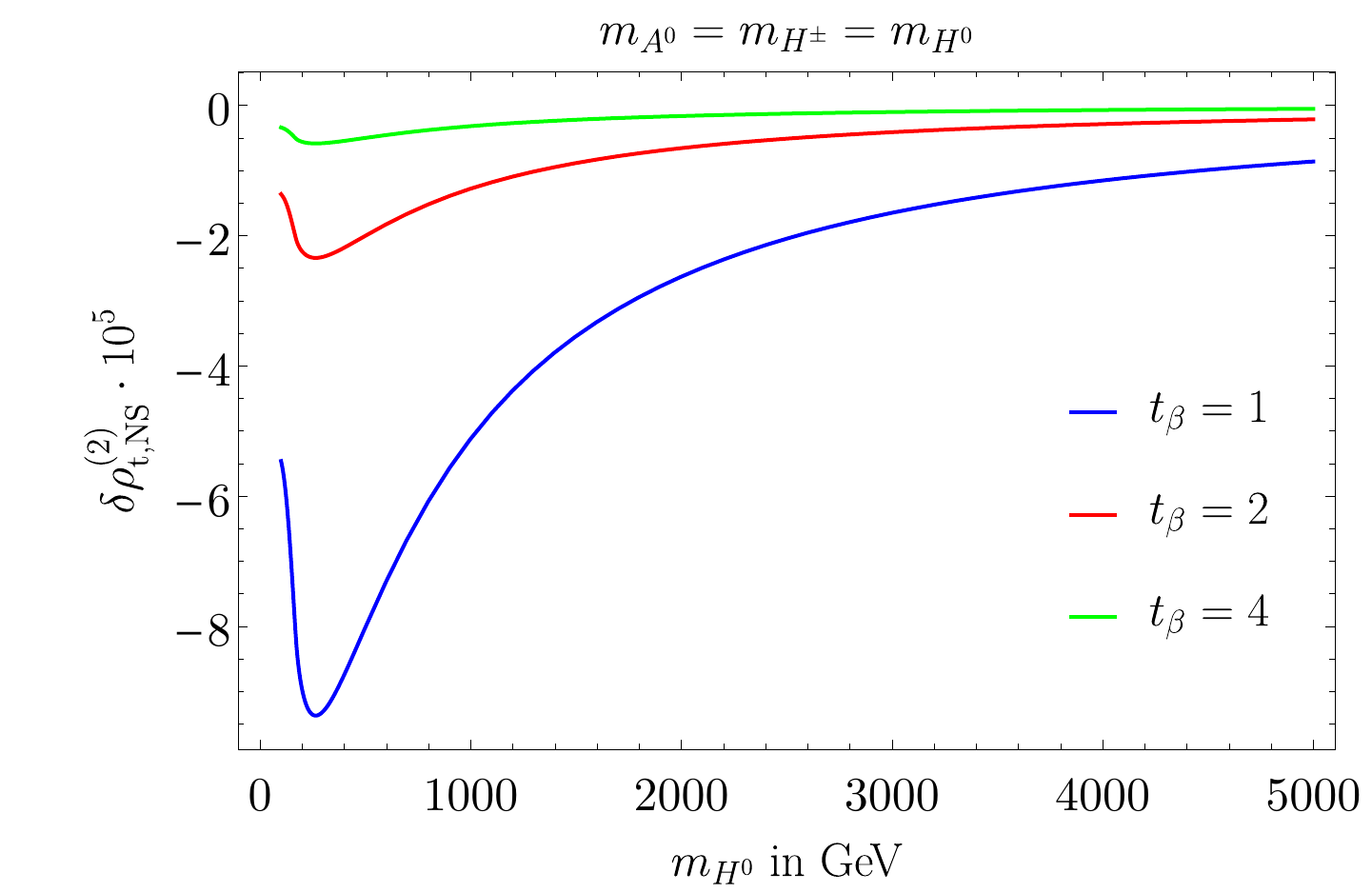}
 \includegraphics[width=\linewidth]{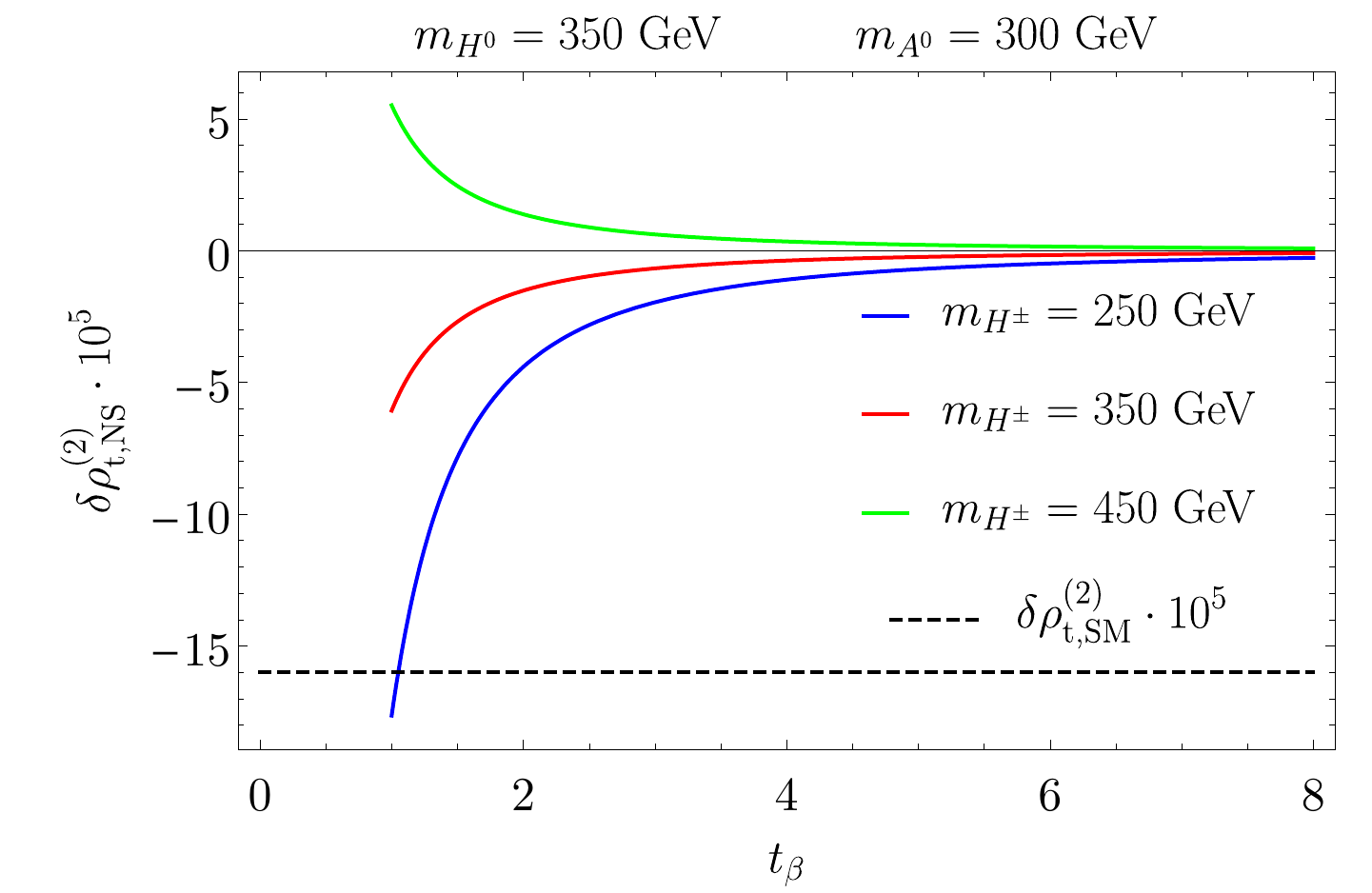}
 \caption{Analysis of $\delrho[t,NS]{2}$. The upper panel presents a variation of the degenerate masses $\MHH$, $\MAO$ and $\MHp$ up to large values. The solid lines correspond to different values of $t_\beta$.
  In the lower panel $\delrho[t,NS]{2}$ is plotted as a function of $t_\beta$ for different values of $m_H^\pm$. The masses of $H^0$ and $A^0$ are fixed at $\MHH=350\text{ GeV}$ and $\MAO=300\text{ GeV}$. The value of the two-loop top-Yukawa correction in the SM, $\delta\rho^{(2)}_{\rm t,SM} = -1.60\cdot 10^{-4}$, is shown by the black dashed line for comparison.}
 \label{Fig:NStopResults}
 \end{center}
\end{figure}

\noindent
We start with the analysis of the contribution $\delrho[t,NS]{2}$ which is originating from the coupling between the top quark and the non-standard scalars. As a first test of our result we examine the behaviour in the so-called decoupling limit \cite{Gunion:2002zf}, in which the masses of the non-standard scalars are much larger than $m_h^0$. In this limit the scalar sector of the THDM can be described by an effective theory which  is identical to the SM Higgs sector. Consequently we expect $\delrho[t,NS]{2}$ to vanish for large, equal non-standard Higgs masses. The decoupling scenario is investigated in the upper panel of figure~\ref{Fig:NStopResults}, where $\delrho[t,NS]{2}$ is shown for degenerate masses of the non-standard scalars. The solid lines represent results for different values of $t_\beta$. Since the top-Yukawa coupling breaks the custodial symmetry this contribution is still non-zero, even if the custodial symmetry in the Higgs potential is restored by equal masses of the charged and neutral Higgs states. As expected it approaches zero when the masses increase. Moreover, we can see that larger values of $t_\beta$ suppress the correction. The reason is that the coupling of the top quark to the scalars $H^0$, $A^0$ and $H^\pm$ scales with $t_\beta^{-1}$ in the alignment limit (see section~\ref{subsec:Align}).

The influence of $t_\beta$ is visualised on the lower panel of figure~\ref{Fig:NStopResults} with $\delrho[t,NS]{2}$ for the mass configurations as described by the legend, showing the decrease of the contribution with $t_\beta$. In addition different mass splittings between charged and neutral scalars yield noticable deviations in the result and can even lead to different signs.
In general, the top-Yukawa contribution is of the order of the SM  value $\delta\rho^{(2)}_{\rm t,SM}$ 
 or smaller.

In order to test the validity of the top-Yukawa approximation, we repeated our calculation including also the contribution from the bottom-Yukawa coupling.  
In the THDM of type-I and type-X the additional corrections from the bottom-Yukawa coupling are
negligibly small, as expected from their suppression by the $b$-quark mass
(see section \ref{Sec:Outline}). 
In the type-II and type-Y models, the contribution from the bottom-Yukawa coupling can be enhanced for large values of $\TB$ since the coupling of the $b$-quark to the non-standard scalars carries a factor $\TB$ in the alignment limit. 
Additional two-loop contributions from finite $m_b$ that reach the level of
$\delrho[t,SM]{2}$,  require  $\TB\simeq 40-50$. 
For such large values of $\TB$, however, one has to prevent the non-standard scalar
self-couplings from becoming non-perturbative by
restricting the parameter $\lfive$ to be very close to 
$\lfive v^2= 2\MHH^2$\cite{Akeroyd:2000wc,Das:2015qva}. 
Moreover,  the constraints from flavour physics give further significant restrictions 
for large values of $\TB$ 
(see for example \cite{Deschamps:2009rh,Enomoto:2015wbn}).

\subsection{Results for the non-standard scalar contribution}
\label{Sec:NSResults}

\begin{figure}[ht]
 \begin{center}
  \includegraphics[width=\linewidth]{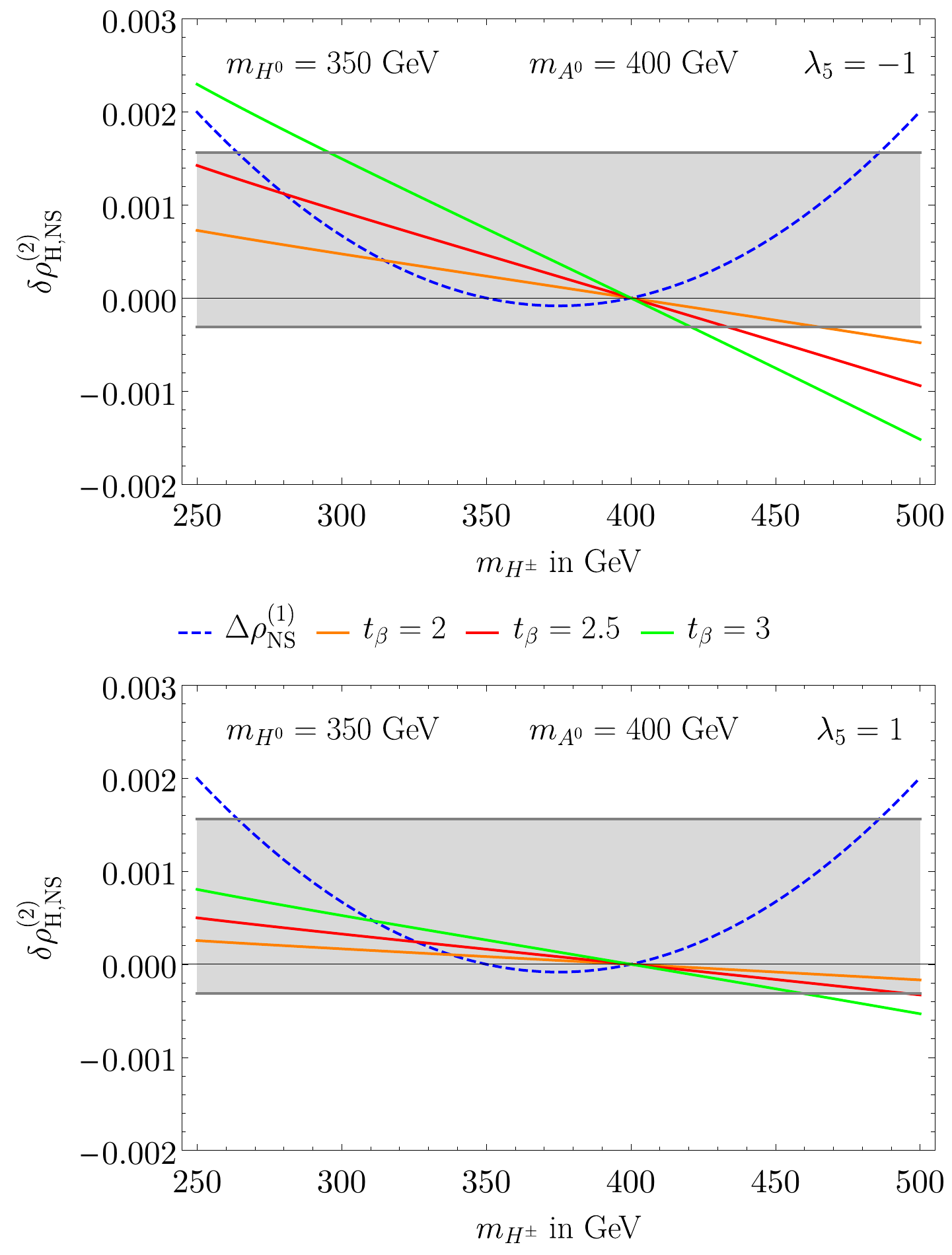}
 \caption{Effect of mass differences between neutral and charged scalars on $\delrho[H,NS]{2}$ for $\lambda_5=\pm1$. The neutral masses are fixed at $\MHH=350\text{ GeV}$ and $\MAO=400\text{ GeV}$. The mass of $H^\pm$ is varied from $250\text{ GeV}$ to $500\text{ GeV}$. The solid lines represent different values of $t_\beta$ as explained in the legend. The blue dashed line shows the non-standard one-loop correction $\Delrho[NS]{1}$ for comparison. The grey area depicts the bounds from the experimental limits of the $T$ parameter.} 
 \label{Fig:drhoNSHmasssplit}
 \end{center}
\end{figure}

\begin{figure}[ht]
 \begin{center}
  \includegraphics[width=\linewidth]{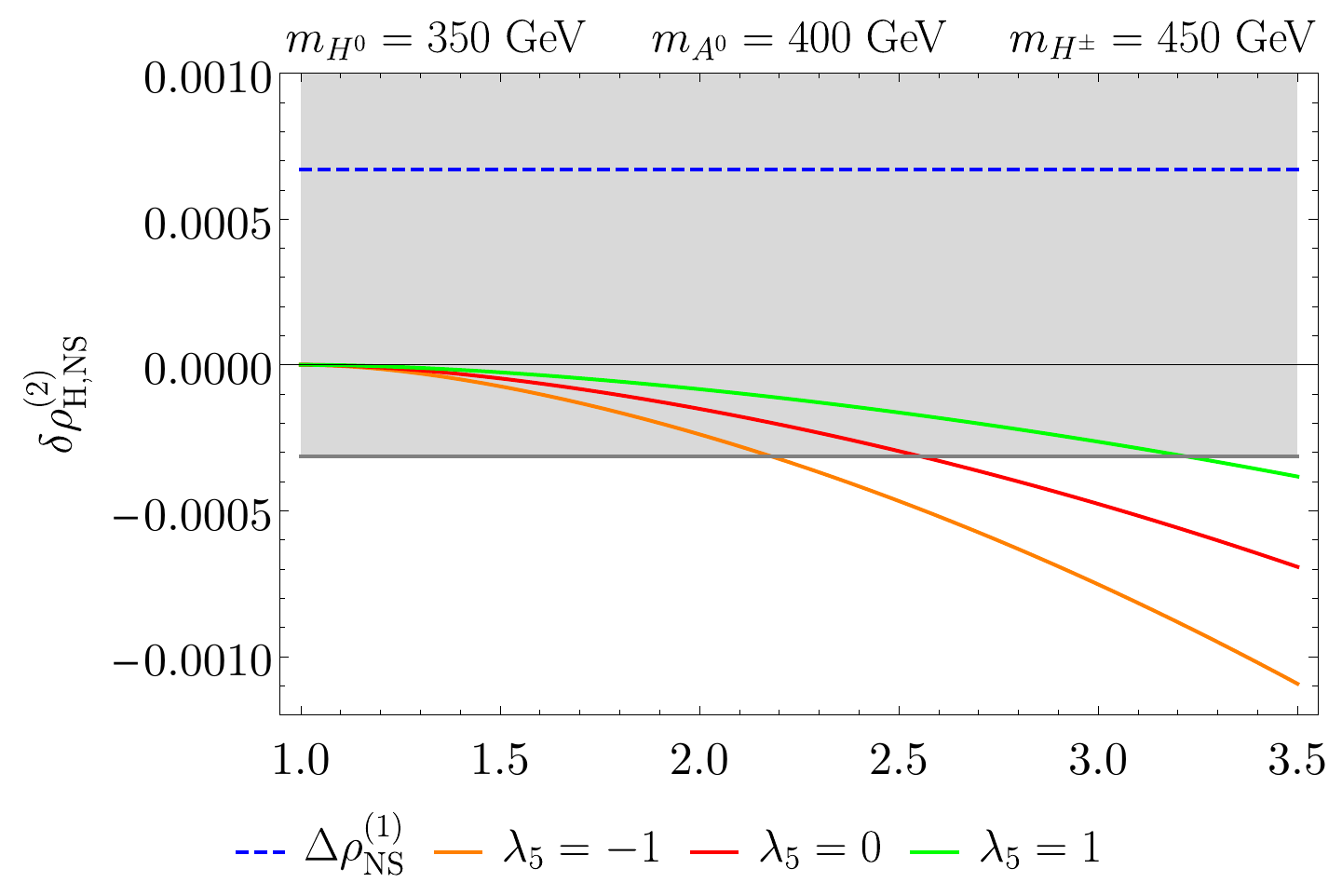}
  \caption{Influence of a variation of $t_\beta$ on $\delrho[H,NS]{2}$ for the specified mass configurations. The result is plotted for different values of $\lambda_5$. The blue dashed line gives the value of the non-standard one-loop correction $\Delrho[NS]{1}$ for the specified masses. The grey area depicts the bounds from the experimental limits of the $T$ parameter.}
 \label{Fig:drhoNSHtbvari}
 \end{center}
\end{figure}

We now discuss the numerical results of the contribution $\delrho[H,NS]{2}$ which originates from the coupling between three non-standard scalars as described in section~\ref{Subsec:drhoNSHiggs}. The influence of a mass splitting between charged and neutral scalars is presented in figure~\ref{Fig:drhoNSHmasssplit}. The two panels show results for $\MHH=350 \text{ GeV}$, $\MAO=400\text{ GeV}$ and $\lambda_5=\pm1$. The variation of $\MHp$ is performed such that it yields similar mass differences for the specified parameter settings. The different lines correspond to different values of $t_\beta$ as defined in the legend. For comparison the blue dashed line displays the result for the one-loop non-standard correction $\Delrho[NS]{1}$. The grey area indicates the bounds from the $T$ parameter in~(\ref{Eq:Texp}). 

We see that the contribution $\delrho[H,NS]{2}$ can give corrections to the $\rho$ parameter which are comparable in size or even larger than the one-loop correction. The reason are the new couplings between three non-standard scalars which enter for the first time in the two-loop contribution. Adding the two-loop corrections to the one-loop result can lead to noticeable modifications of the parameter region allowed by the constraints on $T$.

The triple non-standard scalar couplings arise from the term $V_{IV}$ of the potential in \eqref{Eq:PotAlign}, when the vacuum expectation value
\begin{equation}
 \langle\Phi_\text{SM}\rangle=\frac{1}{\sqrt{2}}
\begin{pmatrix}
 0 \\
 v
\end{pmatrix}
\end{equation}
is inserted for the doublet $\Phi_\text{SM}$. Since they enter quadratically in all the diagrams in figure~\ref{Fig:NSHiggs2Loop}, the contribution $\delrho[H,NS]{2}$ is proportional to (see \eqref{Eq:V_IV})
\begin{equation}
% \frac{1}{\TTwoB^2}\left(\frac{2\MHH^2}{v^2}-\lfive\right)^2\\=
\frac{1}{4}\left(\frac{1}{\TB}-\TB\right)^2\left(\frac{2\MHH^2}{v^2}-\lfive\right)^2.
 \label{Eq:Prefactor}
\end{equation}
The prefactor explains the strong influence of $t_\beta$ on the results in figure~\ref{Fig:drhoNSHmasssplit}. The enhancement of the coupling can be weakened for positive values of $\lambda_5$ (see the lower panel of figure~\ref{Fig:drhoNSHmasssplit}) or increased for negative values of  $\lambda_5$ (see the upper panel of figure~\ref{Fig:drhoNSHmasssplit}).

The effect of the custodial transformations described in section~\ref{Sec:CustSym} is also visible in figure~\ref{Fig:drhoNSHmasssplit}. The one-loop contribution $\Delrho[NS]{1}$  is zero for $\MHH=\MHp$ and $\MAO=\MHp$ since it originates only from the part $V_{III}$ of the potential which is custodial symmetric for these two mass settings.  
As explained in section~\ref{Sec:CustTrans1} the part $V_{IV}$ is
invariant under the custodial transformation for
$\chi=0$. Consequently $\delrho[H,NS]{2}=0$ for $\MAO=\MHp$ since all
the involved couplings are custodial invariant for this mass degeneracy. 
However, for $\MHH=\MHp$ we have $\delrho[H,NS]{2}\neq 0$ since in  that case  $V_{III}$ 
is invariant only under custodial transformations for $\chi=\frac{\pi}{2}$,
but then $V_{IV}$ is not invariant and
the triple couplings between three non-standard scalars 
hence break the custodial symmetry 
(see section~\ref{Sec:CustTrans2}).

The dependence of $\delrho[H,NS]{2}$ on $t_\beta$ is visualized 
directly in figure~\ref{Fig:drhoNSHtbvari} for different values of $\lfive$, 
displaying the increase with $t_\beta$ 
and the modification by the choice of $\lambda_5$ according to~\eqref{Eq:Prefactor}.

\subsection{Results for the mixed scalar contribution}

\begin{figure}[ht]
 \begin{center}
	\includegraphics[width=\linewidth]{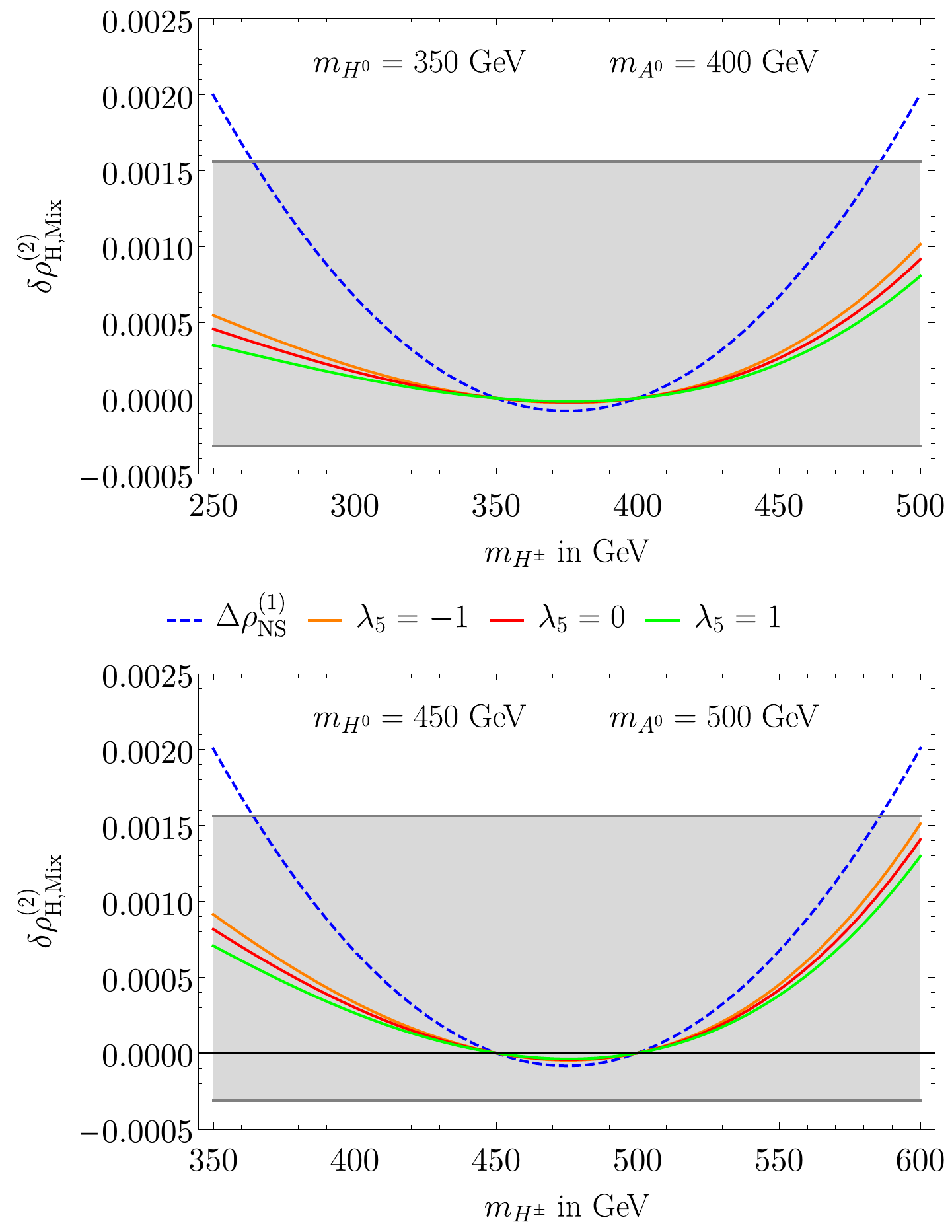}
 \caption{Influence of  mass splitting between charged and neutral
   scalars on $\delrho[H,Mix]{2}$. The two plots show different values
   of $m_{H^0}$ and $m_{A^0}$, and the variation of $\MHp$ leads to
   comparable mass differences for the different mass
   configurations. The results are independent of $t_\beta$. 
  The different lines represent different values of $\lambda_5$. 
  The blue dashed line shows the result of the non-standard one-loop
  correction $\Delrho[NS]{1}$ for comparison. 
 The grey area depicts the bounds from the experimental limits of the $T$ parameter.}
 \label{Fig:drhomixsplit}
\end{center}
\end{figure}
In the last part we discuss the contribution $\delrho[H,Mix]{2}$ from the interaction of the SM-like scalars $h^0$, $G^0$, $G^\pm$ with the non-standard scalars $H^0$, $A^0$, $H^\pm$. Similar to the one-loop correction $\Delrho[NS]{1}$ it originates only from the part $V_{III}$ of the potential in \eqref{Eq:PotAlign}. Consequently it is independent of $t_\beta$ (see \eqref{Eq:V_III}). 

In figure~\ref{Fig:drhomixsplit} we analyse the influence of a mass splitting between the charged and neutral scalars. We show two scenarios for different values of $m_{H^0}$ and $m_{A^0}$, while the mass of $m_{H^\pm}$ is varied in such a way that the mass splittings are comparable. The three solid lines present the results for different values of $\lambda_5$. The blue dashed line gives the one-loop contribution $\Delrho[NS]{1}$ for comparison. 

The results of figure~\ref{Fig:drhomixsplit} can again be explained with the help of the custodial symmetry. As discussed in section~\ref{Sec:CustSym} there are the two possible ways,  
\begin{equation}
 \MHH=\MHp
\end{equation}
or
\begin{equation}
 \MAO=\MHp   \, ,
\end{equation}
to restore a custodial symmetry in $V_{III}$. For these two mass configurations $\Delrho[NS]{1}$ and $\delrho[H,Mix]{2}$ vanish, since they do not contain any additional custodial-symmetry breaking couplings.

While the one-loop contribution originates only from the coupling of the non-standard scalars to the Goldstone bosons, new couplings between $h^0$ and the non-standard scalars enter the two-loop diagrams in figure~\ref{Fig:Mix2Loop}. These are proportional to the combination
\begin{equation}
% \left(
2 m_{S}^2+m_{h^0}^2-\lambda_5 v^2
\label{Eq:h0NScoupling}
%\right)
\end{equation}
where $S$ can be either of $H^0$, $A^0$ or $H^\pm$, depending on which scalar couples to $h^0$. The effect of these new couplings is clearly visible  in the numerical results. By comparing the upper and the lower panel of figure~\ref{Fig:drhomixsplit} we see that larger masses of the non-standard scalars yield larger values of $\delrho[H,Mix]{2}$. 
In addition the couplings can be enhanced or suppressed by negative or positive values of $\lambda_5$, 
which explains the variation between the different solid lines representing different values of $\lambda_5$.

Since the correction $\delrho[H,Mix]{2}$ is independent of $t_\beta$ it will be the dominant scalar two-loop correction to the $\rho$ parameter for $t_\beta\approx 1$ where $\delrho[H,NS]{2}$ is small. However, for $m_{H^0}=m_{H^\pm}$ both the one-loop correction $\Delrho[NS]{1}$ and $\delrho[H,Mix]{2}$ vanish independently of $t_\beta$,
and $\delrho[H,NS]{2}$ is the only 
remaining scalar correction to the $\rho$ parameter (for $t_\beta \neq 1$). 

For the Inert-Higgs-Doublet-Model (IHDM),
as explained in section~\ref{Sec:IHDM}, 
the only non-standard two-loop correction to the $\rho$ parameter is equivalent to $\delrho[H,Mix]{2}$. 
Conventionally, the parameter $\mu_2^2$ is often used as a free input parameter. The results in 
fig.~\ref{Fig:drhomixsplit} can easily be interpreted in the IHDM 
by means of the relation~\eqref{Eq:mu2relation} to trade $\lambda_5$ for $\mu_2^2$.

\subsection{Results for a light pseudoscalar}

\begin{figure}[ht]
\includegraphics[width=\linewidth]{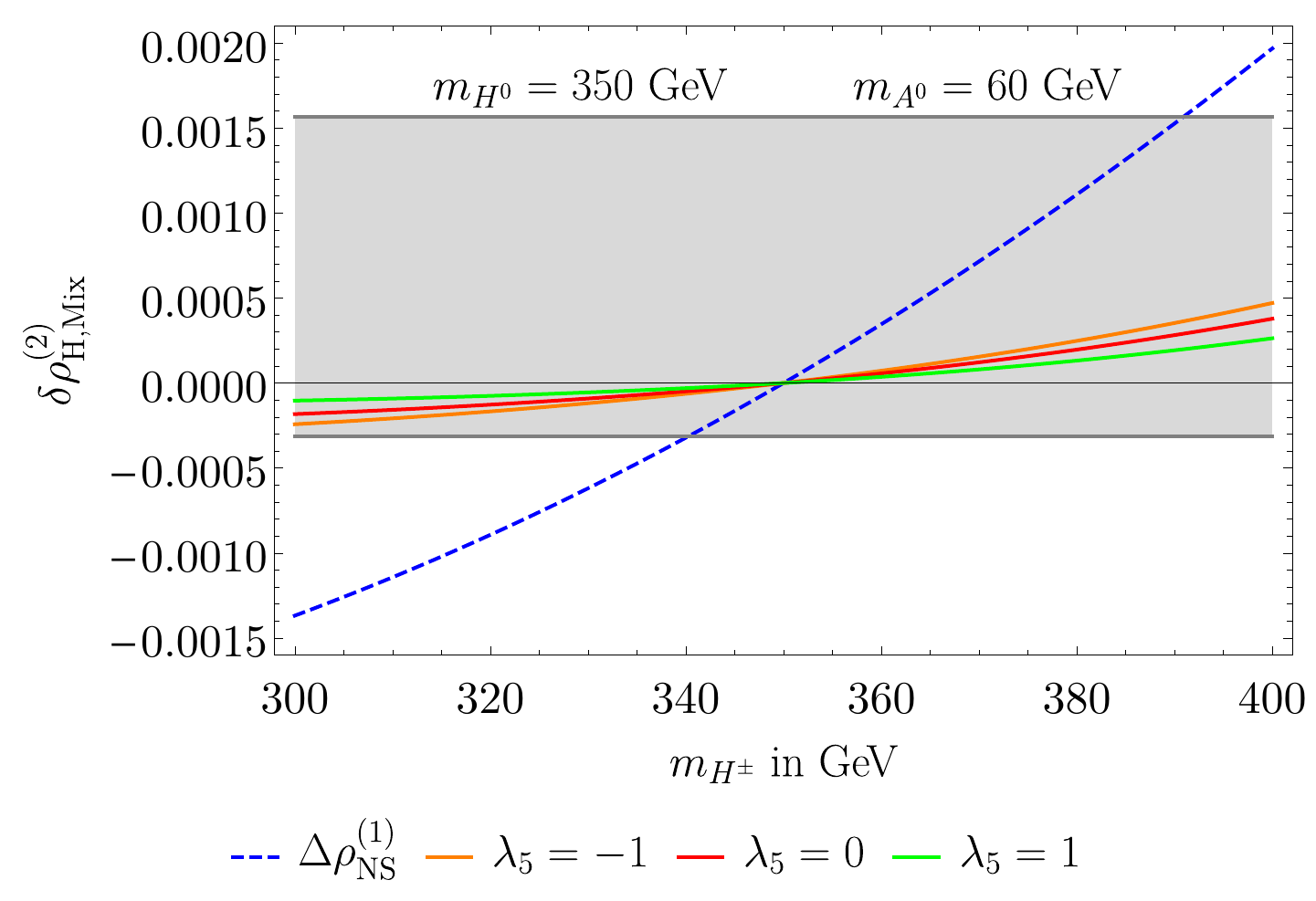}
\caption{
Results for $\delrho[H,Mix]{2}$ for $\MAO=60\text{ GeV}$. The mass of
$H^0$ is fixed at $350\text{ GeV}$. 
The different solid lines correspond to different values of $\lfive$, and the results are independent of $\TB$. The non-standard one-loop correction $\Delrho[NS]{1}$ is shown by the blue dashed line.
The grey area corresponds to the bounds from the experimental limits of the $T$ parameter. 
}
\label{Fig:drhoMixlightA0}
\end{figure}

A light pseudoscalar with $m_{A^0}<125\text{ GeV}$ can still be possible in the THDM (for a detailed analysis see \cite{Bernon:2014nxa}). 
The non-standard top-Yukawa contribution is similar to the case discussed in 
section~\ref{Sec:NStResults}.
For the scalar contributions, a general feature of a light $A^0$  boson  
consists in a large splitting of the two zeros of the dashed line 
in figures~\ref{Fig:drhoNSHmasssplit} and~\ref{Fig:drhomixsplit}.
The area around the zero at $\MHp=m_{A^0}$ is
excluded by the absence of light charged Higgs bosons. Hence, only the other
zero at $\MHp=\MHH$ is phenomenologically acceptable and deserves a closer inspection. 
The one-loop contribution  $\Delrho[NS]{1}$ and the two-loop contribution $\delrho[H,Mix]{2}$
are both independent of~$\TB$; they are displayed in figure~\ref{Fig:drhoMixlightA0} where one can see
that $\delrho[H,Mix]{2}$ follows the direction of $\Delrho[NS]{1}$ and thus amplifies the 
dependence on the mass splitting between $H^0$ and $H^\pm$,
disfavoring the case $m_{H^\pm} < m_{H^0}$. 

\begin{figure}[ht]

%\vspace*{1cm}
\includegraphics[width=\linewidth]{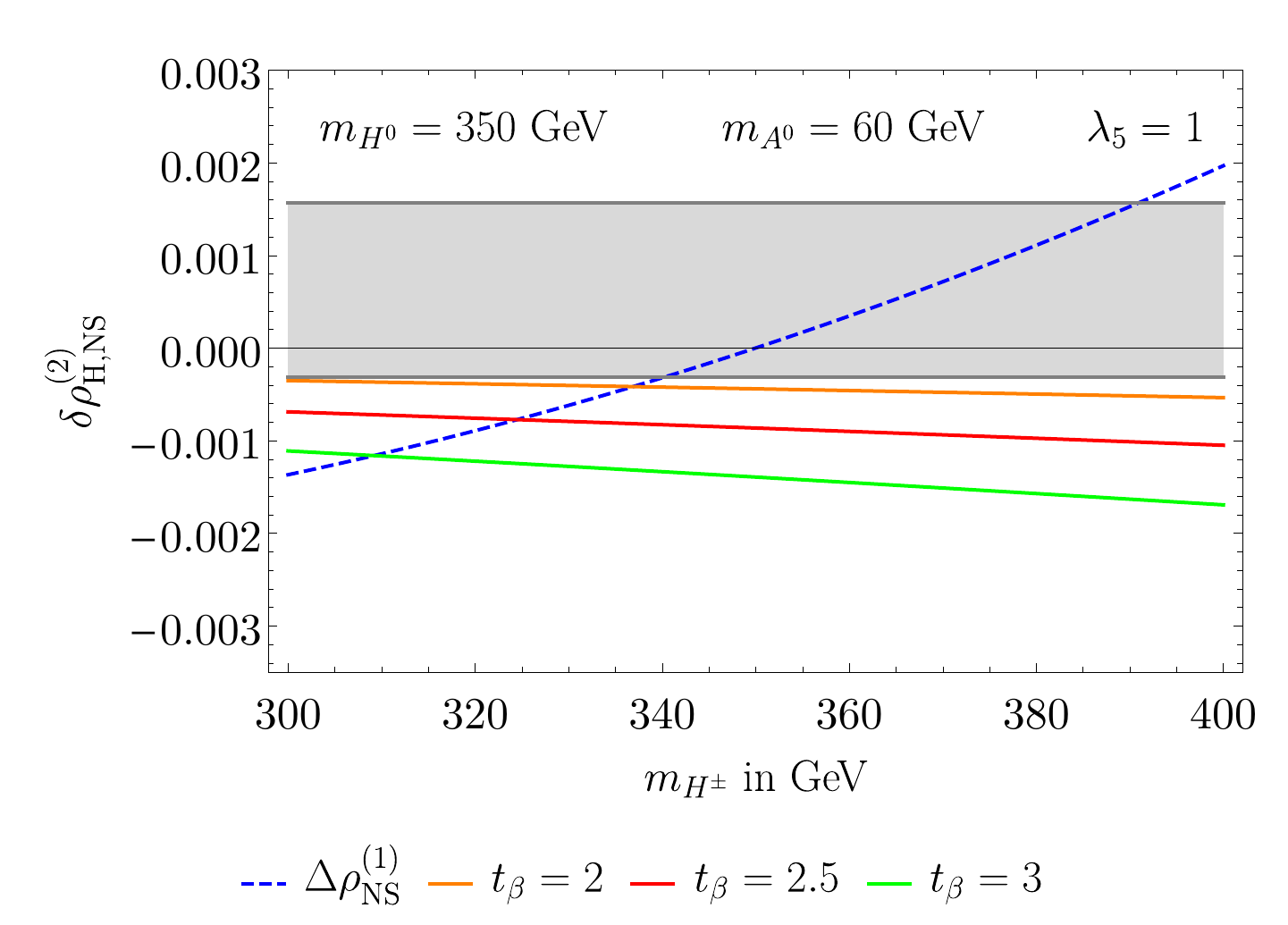}
\caption{
Results for $\delrho[H,NS]{2}$ for $\MAO=60\text{ GeV}$. The mass of $H^0$ is fixed at $350\text{ GeV}$. 
The different solid lines correspond to different values of $\TB$. The non-standard one-loop correction $\Delrho[NS]{1}$ is shown by the blue dashed line.
The grey area corresponds to the bounds from the experimental limits of the $T$ parameter. 
}
\label{Fig:drhoHNSlightA0}
\end{figure}

The purely non-standard scalar contribution $\delrho[H,NS]{2}$ vanishes for $\TB=1$, but otherwise has a strong
variation with~$\TB$ (and $\lfive$). It is shown in figure~\ref{Fig:drhoHNSlightA0}, the analogous plot 
to figure~\ref{Fig:drhoNSHmasssplit}, now with a light $A^0$. Since the common zero  of all curves
corresponds to $m_{A^0}$, the two-loop contribution $\delrho[H,NS]{2}$ is always negative for $m_{H^0, H^\pm} > m_{A^0}$ 
and thus can diminish  $\Delrho[NS]{1}$ substantially for $m_{H^\pm} > m_{H^0}$ when $\TB$ increases.
Again, the situation $m_{H^\pm} < m_{H^0}$ is disfavored.

For $\MAO<\MhO/2$, the coupling of $h^0$ to two pseudoscalars
has to be small to suppress the decay channel 
$h^0\rightarrow A^0 A^0$ \cite{Bernon:2014nxa}. 
In the alignment limit this requires to restrict the value of $\lfive$
to  $\lfive v^2 \simeq 2\MAO^2+\MhO^2$ 
(see \eqref{Eq:h0NScoupling}).

Scenarios with a light $A^0$ are especially interesting in the THDM,
since Barr-Zee type two-loop diagrams can provide an explanation for
the $3\sigma$ difference between the SM prediction and the measured
value of the muon anomalous magnetic moment $a_\mu$
\cite{Chang:2000ii}. An improved agreement between theory and
experiment consistent with several theoretical and experimental
constraints can be achieved in a type-X model with very large values of $\TB$
(see \cite{Broggio:2014mna,Chun:2016hzs} and references
therein). Usually $\MHp=\MHH$ is assumed, to fulfill the constraints
from electroweak precision observables. For the $\rho$ parameter this means
vanishing contributions from $\Delrho[NS]{1}$ and
$\delrho[H,Mix]{2}$. Furthermore, the top-Yukawa contribution
$\delrho[t,NS]{2}$ is strongly suppressed.
However, for such large values of $\TB$, the non-standard scalar contribution
$\delrho[H,NS]{2}$ would completely run out of control unless the
scalar self-coupling is kept small by adjusting $\lfive$ very close  to
$\lambda_5 =2\MHH^2/v^2$. 
An additional aspect of type-X models with very large $\TB$ is the enhanced  
Yukawa coupling of the $\tau$ lepton. This could yield a further
two-loop contribution to the $\rho$ parameter, which we  did not consider in this work.

\section{Conclusions}
\label{Sec:conclusions}

We have given an overview over the calculation of the two-loop contributions
to the $\rho$-parameter in the $CP$-conserving Two-Higgs-Doublet Model
where one of the $CP$-even scalars ($h^0$) is identified with the scalar resonance
at 125 GeV observed by the LHC experiments ATLAS and CMS.
The approximation of the gauge-less limit and massless fermions except the top quark
yield the leading contributions from the top-Yukawa coupling and the self-couplings of the
Higgs bosons, which can be separated into standard and non-standard contributions.
As already at the one-loop level, the non-standard contributions from the scalar self-interactions 
are  particularly sensitive to mass splittings between neutral and charged scalars.
As a new feature, the two-loop contributions have a significant dependence on the 
parameters $\tan\beta$ and $\lambda_5$, the coefficient of the THDM scalar potential 
that is not fixed by the masses of the neutral and charged Higgs bosons, and thus can
modify the one-loop  result substantially. Moreover, this significant dependence on 
the additional parameters can be exploited to get more indirect information on the
Higgs potential from electroweak precision data than with the 
currently available one-loop calculations. 
 
The loop correction $\Delta\rho$ to the $\rho$-parameter is an important entry 
in the calculation of electroweak precision observables, parametrizing dominant
universal contributions from particles with mass splitting in isospin doublets,
in the THDM in particular from
%in the SM and extensions with $\rho=1$ at the tree-level.
neutral and charged Higgs bosons. 
For an estimate of the impact of a shift in $\Delta\rho$
on the prediction of the $W$ mass and the effective weak mixing angle 
$\sin^2 \theta_{\rm eff}$ at~$M_Z$, one can use the approximate expressions 
\begin{eqnarray}
\Delta M_W \simeq \frac{M_W}{2} \, \frac{c_W^2}{c_W^2-s_W^2} \, \Delta\rho, \\
\Delta \sin^2 \theta_{\rm eff} \simeq -  \frac{c_W^2 s_W^2}{c_W^2-s_W^2} \, \Delta\rho ,
\end{eqnarray}
to translate the two-loop contribution to $\Delta\rho$ from the non-standard Higgs  sector
obtained in this paper into shifts of the observables. 
An accurate evaluation of the precision observables and implications from comparisons with  
experimental data requires a more detailed study, which will be presented in a forthcoming publication.

\begin{acknowledgements}
This work was supported in part by the Deutsche Forschungsgemeinschaft (DFG)
under Grant No.~EXC-153
(Excellence Cluster {\it Structure and Origin of the Universe}).
We thank Georg Weiglein for useful discussions and Thomas Hahn and Sebastian Pa\ss ehr for
their helpful support in the installation and handling of the two-loop calculational tools.
\end{acknowledgements}

%\clearpage

\begin{appendix}

\section{Feynman rules for the counterterm vertices}
\label{App:Feynmanrules}

In the counterterm vertices in the diagrams in figure~\ref{Fig:VV_t_CTDiagramme} and figure~\ref{Fig:VV_NS_CTDiagramme} we keep only the counterterms which survive the gauge-less limit.  These are the mass counterterms of the top quark and the non-standard scalars and the renormalization constant $\dSWsq$ which has a remaining contribution in the gauge-less limit (see section~\ref{Sec:Renormalization}).  All field counterterms are dropped since they either cancel in the full result or vanish in the gauge-less limit. In the vertices all the momenta are considered as incoming.
Dropping field renormalization, the scalar--scalar two-point vertex counterterm takes the form 
\begin{center}
\parbox[c]{0.55\linewidth}{
 \includegraphics[width=0.25\textwidth]{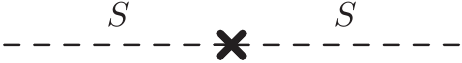}
 }
 \parbox[c]{0.35\linewidth}{\vspace{2.5mm}$=-i\delta m_{S}^2$}
\end{center}
 for  $S=h^0,H^0,A^0,H^\pm,G^0,G^\pm$.
 
The Feynman rule involving the top-quark mass counterterm is given by
\begin{center}
 \parbox[c]{0.55\linewidth}{
 \includegraphics[width=0.25\textwidth]{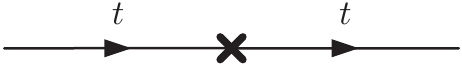}\vspace{2.5mm}
 }
 \parbox[c]{0.35\linewidth}{\vspace{2.5mm}$=-i\delta m_t$.\vspace{2.5mm}}
\end{center}
The renormalization conditions for the mass counterterms are given in section~\ref{Sec:Renormalization}.

For the coupling of a massive gauge boson to two scalars we obtain the following counterterm vertices: 
 \begin{center}
\fbox{
\parbox[c]{0.45\linewidth}{
\includegraphics[width=0.25\textwidth]{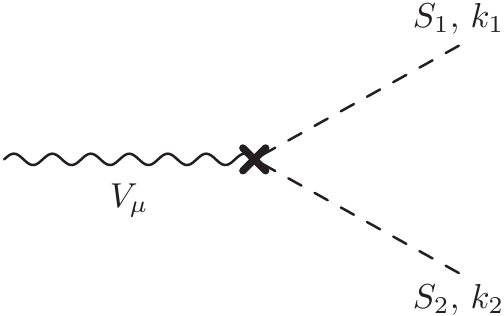}
}
\parbox[c]{0.5\linewidth}{
$=\left(k_{1}-k_{2}\right)_\mu C \left[V_\mu,S_1,S_2\right]$
}
}
\begin{footnotesize}
 \begin{align}
  C[Z_\mu,h^0,G^0]&=\frac{e}{4 s_W c_W}\frac{\left(s_W^2-c_W^2\right)}{c_W^2}\frac{\dSWsq}{s_W^2},\notag\\
  C[Z_\mu,H^0,A^0]&=-\frac{e}{4 s_W c_W}\frac{\left(s_W^2-c_W^2\right)}{c_W^2}\frac{\dSWsq}{s_W^2},\notag\\
  C[Z_\mu,H^-,H^+]&=-i\frac{e}{4 s_W c_W^3}\frac{\dSWsq}{s_W^2},\notag\\
  C[Z_\mu,G^-,G^+]&=-i\frac{e}{4 s_W c_W^3}\frac{\dSWsq}{s_W^2},\notag\\
  C[W_\mu^\pm,h^0,G^\mp]&=\pm i\frac{e}{4 s_W }\frac{\dSWsq}{s_W^2},\notag\\
  C[W_\mu^\pm,H^0,H^\mp]&=\mp i\frac{e}{4 s_W }\frac{\dSWsq}{s_W^2},\notag\\
  C[W_\mu^\pm,A^0,H^\mp]&=-\frac{e}{4 s_W }\frac{\dSWsq}{s_W^2},\notag\\
  C[W_\mu^\pm,G^0,G^\mp]&=-\frac{e}{4 s_W }\frac{\dSWsq}{s_W^2},\notag
 \end{align}
\end{footnotesize}
\end{center}
\newpage
For the coupling of two massive gauge bosons to two scalars the  counterterm vertices are 
\begin{center}
\fbox{
\parbox[c]{0.45\linewidth}{
\includegraphics[width=0.25\textwidth]{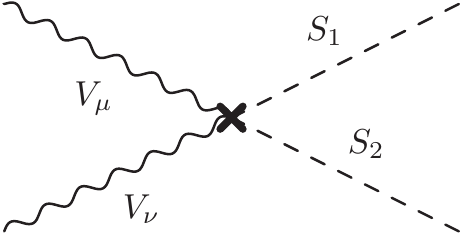}
}
\parbox[c]{0.45\linewidth}{
$= g_{\mu\nu} C \left[V_\mu,V_\nu,S_1,S_2\right]$
}
}
\begin{footnotesize}
 \begin{align}
  C[Z_\mu,Z_\nu,h^0,h^0]&=i\frac{e^2}{2 s_W^2 c_W^2}\frac{s_W^2-c_W^2}{c_W^2}\frac{\dSWsq}{s_W^2},\notag\\
  C[Z_\mu,Z_\nu,H^0,H^0]&=i\frac{e^2}{2 s_W^2 c_W^2}\frac{s_W^2-c_W^2}{c_W^2}\frac{\dSWsq}{s_W^2},\notag\\
  C[Z_\mu,Z_\nu,A^0,A^0]&=i\frac{e^2}{2 s_W^2 c_W^2}\frac{s_W^2-c_W^2}{c_W^2}\frac{\dSWsq}{s_W^2},\notag\\
  C[Z_\mu,Z_\nu,H^+,H^-]&=i\frac{e^2}{2 s_W^2 c_W^2}\frac{s_W^2-c_W^2}{c_W^2}\frac{\dSWsq}{s_W^2},\notag\\
  C[Z_\mu,Z_\nu,G^0,G^0]&=i\frac{e^2}{2 s_W^2 c_W^2}\frac{s_W^2-c_W^2}{c_W^2}\frac{\dSWsq}{s_W^2},\notag\\
  C[Z_\mu,Z_\nu,G^+,G^-]&=i\frac{e^2}{2 s_W^2 c_W^2}\frac{s_W^2-c_W^2}{c_W^2}\frac{\dSWsq}{s_W^2},\notag\\
  C[W_\mu^+,W_\nu^-,h^0,h^0]&=-i\frac{e^2}{2 s_W^2}\frac{\dSWsq}{s_W^2},\notag\\
  C[W_\mu^+,W_\nu^-,H^0,H^0]&=-i\frac{e^2}{2 s_W^2}\frac{\dSWsq}{s_W^2},\notag\\
  C[W_\mu^+,W_\nu^-,A^0,A^0]&=-i\frac{e^2}{2 s_W^2}\frac{\dSWsq}{s_W^2},\notag\\
  C[W_\mu^+,W_\nu^-,H^+,H^-]&=-i\frac{e^2}{2 s_W^2}\frac{\dSWsq}{s_W^2}.\notag\\
  C[W_\mu^+,W_\nu^-,G^0,G^0]&=-i\frac{e^2}{2 s_W^2}\frac{\dSWsq}{s_W^2},\notag\\
  C[W_\mu^+,W_\nu^-,G^+,G^-]&=-i\frac{e^2}{2 s_W^2}\frac{\dSWsq}{s_W^2},\notag 
 \end{align}
\end{footnotesize}
\end{center}

The counterterm vertices between the gauge bosons and the fermions in figure~\ref{Fig:VV_t_CTDiagramme} are given by
\begin{center}
\fbox{
\parbox[c]{0.45\linewidth}{
\includegraphics[width=0.25\textwidth]{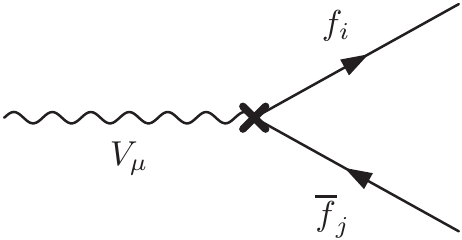}
}
\parbox[c]{0.45\linewidth}{
$= \left(C_- \omega_- + C_+ \omega_+\right)\gamma_\mu$,
}
}
\begin{footnotesize}
\begin{align}
&W^{+}\overline{t} b:\quad C_+=0,\quad C_-=i\frac{e}{2\sqrt{2}s_W}\frac{\dSWsq}{s_W^2},\notag\\
&W^{-} \overline{b} t:\quad C_+=0,\quad C_-=i\frac{e}{2\sqrt{2}s_W}\frac{\dSWsq}{s_W^2},\notag\\
&Z \overline{t} t:\quad C_+=i\frac{e}{3c_W^2}\frac{s_W}{c_W}\frac{\dSWsq}{s_W^2}, \quad C_-=\frac{ie}{4c_W s_W}\left(1+\frac{1}{3}\frac{s_W^2}{c_W^2}\right)\frac{\dSWsq}{s_W^2}\notag.
\end{align}
\end{footnotesize}
\end{center}

\section{One- and two-loop integrals}

\subsection{Scalar one-loop integrals}
\label{App:1LIntegrals}

Here  we list all the one- and two-loop scalar integrals that are used in our calculation. 
They are evaluated in dimensional regularization \cite{tHooft:1972fi,Bollini:1972ui,Ashmore:1972uj} 
with dimension $D$ of the integrated momentum and the associated mass parameter $\mu_D$,
\begin{equation}
\int d^4q\rightarrow\mu^{4-D}_D\int d^Dq \, .
\end{equation}
The scalar integrals are expanded in $\delta=(D-4)/2$ and the divergencies appear as poles in $\delta$.
 
The reduction of the one-loop tensor integrals to scalar integrals and
their classification is following the work of
\cite{Passarino:1978jh,tHooft:1978xw} (for more details and notation see \cite{Denner:1991kt}).
The only one-loop integrals which are needed for the evaluation of the self-energies are
\begin{equation}
 A_0(m^2)=\int\frac{d^Dq}{i\pi^2} \frac{(2\pi \mu_D)^{(4-D)}}{(q^2-m^2+i\epsilon)},
\end{equation}
\begin{multline}
 B_0(p^2,m_1^2,m_2^2)\\
  =\int \frac{d^Dq}{i\pi^2}\frac{(2\pi \mu_D)^{(4-D)}}{(q^2-m_1^2+i\epsilon)((p+q)^2-m_2^2+i\epsilon)}.
\end{multline}
We need an expansion up to order $\delta$ of the scalar integrals; analytic expressions can be found in \cite{Nierste:1992wg,Hollik:2014bua}.

\subsection{Scalar two-loop integrals}
\label{App:2LIntegrals}
The notation of the two-loop integrals follows the conventions of \cite{Weiglein:1993hd}.
For vanishing external momentum %the techniques 
all two-loop integrals in the self energies can be reduced to the scalar integral

\begin{multline}
 T_{134}\left(m_1^2,m_2^2,m_3^2\right)=\left(\frac{(2\pi \mu_D)^{(4-D)}}{i\pi^2}\right)^2\cdot\\
  \int \frac{ \text{d}^Dq_1 \text{d}^Dq_2}{\left(k_1^2-m_1^2+i \epsilon\right)\left(k_3^2-m_2^2+i \epsilon\right)\left(k_4^2-m_3^2+i \epsilon\right)}
\end{multline}
with $k_1=q_1$, $k_3=q_2-q_1$ and $k_4=q_2$.

This integral can be calculated analytically and the result can be found in \cite{Berends:1994ed,Davydychev:1992mt}.

\end{appendix}

%\clearpage

\bibliographystyle{spphys}
\bibliography{rhopaper}

\end{document}